\documentclass[superscriptaddress,twocolumn,secnumarabic,floatfix,nobibnotes,aps,prd,nofootinbib]{revtex4-2}

\usepackage{graphicx}
\usepackage{bm}
\usepackage{amsmath}
\usepackage{amsfonts}
\usepackage{amssymb}
\usepackage{physics}
\usepackage{microtype}
\usepackage{verbatim}
\usepackage{comment}
\usepackage{soul}
\usepackage[normalem]{ulem}
\usepackage{enumitem}   
\usepackage{multirow}
\usepackage{subfigure}
\usepackage{lipsum}

\usepackage{booktabs}
\usepackage{siunitx}
\usepackage{adjustbox}
\usepackage{threeparttable}

\usepackage{fontawesome5}
\usepackage{tikz}
\usepackage{enumitem}

\usepackage[dvipsnames]{xcolor}
\usepackage[colorlinks = true,
            linkcolor = blue,
            urlcolor  = blue,
            citecolor = blue,
            anchorcolor = blue]{hyperref}
\usepackage{orcidlink} 

\usepackage[capitalize]{cleveref}

\newcommand{\tophattails}{%
    \begin{tikzpicture}[scale=0.15,baseline={(0,-0.1)}]
        \draw[thick] (-1,-0.5) -- (0,-0.5);
        \draw[thick] (2,-0.5) -- (3,-0.5);
        \draw[thick] (0,-0.5) -- (0, 0.5) -- (2,0.5) -- (2,-0.5);
    \end{tikzpicture}%
}

\begin{document}

\title{Early against Late: \\ A contrast on dark energy in light of DESI DR2.}


\author{Miguel A. Zapata\,\orcidlink{0009-0004-8575-4844}} 
\email{miguel\_delacruz@icf.unam.mx}
\affiliation{Instituto de Ciencias F\'{\i}sicas, Universidad Nacional Aut\'onoma de M\'exico, Av. Universidad s/n, Cuernavaca, Morelos, 62210, M\'exico.}

\author{Karim Carrion\,\orcidlink{0000-0002-1798-7978}} 
\email{kcarrion@fisica.unam.mx}
\affiliation{Instituto de F\'isica, Universidad Nacional Aut\'onoma de M\'exico, Circuito de la Investigaci\'on Cient\'ifica, Ciudad Universitaria, Cd. de M\'exico C. P. 04510, M\'exico.} 
\affiliation{Instituto de Ciencias F\'{\i}sicas, Universidad Nacional Aut\'onoma de M\'exico, Av. Universidad s/n, Cuernavaca, Morelos, 62210, M\'exico.}

\author{Gabriela Garcia-Arroyo\,\orcidlink{0000-0002-0599-7036}} 
\email{arroyo@icf.unam.mx}
\affiliation{Instituto de Ciencias F\'{\i}sicas, Universidad Nacional Aut\'onoma de M\'exico, Av. Universidad s/n, Cuernavaca, Morelos, 62210, M\'exico.}

\date{\today}

\begin{abstract}
Recent findings from Dark Energy Spectroscopic Instrument (DESI) Baryon Acoustic Oscillation (BAO) measurements, combined with Type Ia supernovae and Cosmic Microwave Background (CMB) data, suggest potential parameter-level deviations from $\Lambda\text{CDM}$. However, parameter exclusions omit prior-volume penalties, whereas Bayesian evidence uncovers a stark dichotomy between early- and late-time dynamics. To quantify this effect, we perform a Bayesian model comparison that explicitly accounts for the cosmological epoch of dark energy dynamics, contrasting two schemes acting in opposite epochs of cosmic history: pre-recombination Early Dark Energy (EDE) and late-time Chevallier--Polarski--Linder (CPL), both measured against a baseline $\Lambda$CDM. We validate the learned harmonic-mean estimator against \texttt{UltraNest} (at background level) before applying it to the CMB analysis. Background-only data already disfavor both extensions, yielding {$\ln B_{\Lambda\text{CDM},\text{EDE}} = 1.48 \pm 0.18$ and $\ln B_{\Lambda\text{CDM},\text{CPL}} = 2.36 \pm 0.24$}. Including CMB sharpens this result: {EDE is very strongly rejected ($\ln B_{\Lambda\text{CDM},\text{EDE}} \approx 12.61 \pm 0.17$)} with its early-time fraction constrained to $10^3\,\Omega_e^{\rm EDE} = 2.0^{+0.6}_{-1.0}$, whereas CPL remains only mildly disfavored ($\ln B_{\Lambda\text{CDM},\text{CPL}} \approx 2.34 \pm 0.18$), even though 2D parameter posteriors for both models display a multi-$\sigma$ deviation from $\Lambda\text{CDM}$. Thus, Bayesian evidence weakens reported preferences for dark energy dynamics, with the impact depending on the cosmic epoch involved, confirmed by our functional reconstructions of $w_{\rm de}(z)$ and $\Omega_{\rm de}(z)$ from posterior samples. 
\end{abstract}

\maketitle

\section{Introduction}
\label{Introduction}

The prevailing $\Lambda$CDM model, while providing an excellent overall fit to a vast array of cosmological observations \cite{Peebles:2024txt}, is currently confronting anomalies that challenge its completeness. In particular, the recent Data Release 2 (DR2) from the Dark Energy Spectroscopic Instrument (DESI) provides high-precision baryon acoustic oscillation (BAO) measurements \cite{DESI:2025zpo,DESI:2025zgx}. When combined with complementary probes \cite{DESI:2025gwf}, these measurements have intensified the debate around the nature of dark energy, suggesting a potential deviation from a pure cosmological constant and reporting preliminary evidence for an evolving dark energy component at the $2\sigma$ -- $3\sigma$ \cite{DESI:2025zgx}.

Whether recent cosmological anomalies point to a more complex dark energy sector drives the exploration of physics beyond $\Lambda\mathrm{CDM}$, traditionally motivated by persistent discrepancies between early- and late-time inferences of $H_0$ and $S_8$ \cite{2021CQGra..38o3001D,2023Univ....9...94H,Abdalla:2022yfr}. Theoretical extensions generally follow two complementary avenues depending on when the modification takes effect. A straightforward approach involves late-time dynamical scenarios where the dark energy equation of state (EoS), $w_{\rm de}(z)$, evolves with redshift \cite{2019ApJ...883L...3L,2022PDU....3601037R} while remaining subdominant at high redshift.

Alternatively, the Early Dark Energy (EDE) paradigm posits a temporary dark energy component active prior to recombination \cite{2013PhRvD..87h3009P}. By accelerating the pre-recombination expansion, EDE reduces the sound horizon at the drag epoch, thereby yielding a larger CMB-inferred $H_0$ value that directly addresses the Hubble tension \cite{Poulin:2018cxd, 2023ARNPS..73..153K, 2023PDU....4201348P}.

In many EDE scenarios, this early component becomes relevant for a limited period of time and subsequently dilutes away, while the late-time accelerated expansion is still driven by a cosmological constant. In contrast, the behavior considered in this work corresponds to a unified dark energy description \cite{Albrecht:1999rm}, where the same dark energy sector evolves across cosmic time. Its evolution is characterized by distinct phases: it behaves like radiation in the early universe, transitions to a matter-like EoS during the matter-dominated era, and ultimately mimics a cosmological constant at late times. Current observational constraints on EDE, however, remain sensitive to both the specific model realization and the combination of datasets considered, with some analyses favoring a non-negligible early contribution while others remain consistent with a negligible EDE component \cite{Niedermann:2019olb,2023PhRvD.108d3513H, 2024IJMPD..3330003M, 2025PDU....4801902J, CosmoVerseNetwork:2025alb}.

EDE can be realized through different physical mechanisms, with scalar fields providing one of the most widely studied frameworks \cite{Albrecht:1999rm,Poulin:2018cxd}. In these scenarios, the field dynamics and the choice of its potential determine the evolution of the EDE component. Rather than modeling EDE through scalar fields rolling down specific potentials, in this work we adopt a phenomenological approach based on a parametrized EoS \cite{2006JCAP...06..026D,2013PhRvD..87h3009P}. This strategy allows for a model-agnostic exploration of the EDE phase space, capturing the essential macroscopic features without committing to a specific Lagrangian or potential form \cite{Sabla:2022xzj}. 
Importantly, flexible EoS parametrizations have been shown to effectively map the background dynamics of diverse scalar field potentials \cite{Pantazis:2016nky,Adil:2026kfn,Garcia-Arroyo:2024tqq} thereby preserving the physical significance of the constraints. \\
Moreover, it drastically reduces the computational burden by avoiding the numerical challenges associated with scalar fields, thus enabling an efficient parameter estimation.

On the other hand, late-time dark energy models provide departures from a cosmological constant by modifying the expansion history at low redshifts. Several parametrizations have been proposed to describe this possibility and, although their functional forms may differ slightly, many of them lead to qualitatively similar late-time dynamics when constrained by current data \cite{DESI:2025zgx}. For this reason, we consider the Chevallier--Polarski--Linder (CPL) parametrization (also well-known as $w_0w_a $CDM model) \cite{CHEVALLIER_2001,Linder} as a minimal and representative example of this class of departures, providing a useful benchmark against which early-time modifications can be contrasted.

Thus, the contrasting signatures of \emph{early} versus \emph{late} dark energy intend to explore how the inferred nature of dark energy depends on the cosmological epoch in which the modification is assumed to take place. This comparison is also relevant for neutrino masses, since DESI+CMB relaxes the bound when a varying EoS is assumed instead of a cosmological constant \cite{DESI:2025zgx}. This paper examines this contrast, evaluating the interplay between EDE mechanisms against the late dark energy signatures suggested by current large-scale structure measurements, particularly in light of the high-precision data from DESI DR2.

The structure of the paper is as following: In Sec.~\ref{sec:DE_parametrization}, we introduce the early and late-time dark energy parametrizations considered in this work. Sec.~\ref{sec:DE_perturbations} describes the treatment of dark energy perturbations. In Sec.~\ref{sec:Methods&Data}, we present the datasets and statistical methodology used in the analysis. Sec.~\ref{Results} discusses the resulting constraints and model comparison, while Sec.~\ref{Conclusions} summarizes our conclusions. Supplementary material omitted from the main text is provided in the Appendices.

\section{Dark Energy parametrizations}
\label{sec:DE_parametrization}

We focus on two dark energy parametrizations  to test distinct asymptotic  departures from a cosmological constant in the early universe. On the one hand, the EDE scenario considered here allows $w>-1/3$ at early times, approaching matter- or radiation-like behavior, while recovering the accelerated-expansion regime at late times. On the other hand, late-time dark energy remains in the $w<-1/3$ regime throughout cosmic evolution. Together, they allow us to assess, within the same datasets and statistical framework, which type of departure from $\Lambda$CDM is favored by current observations.

Both scenarios enter the background dynamics through the evolution of the dark energy density. For a general evolving $w_{\rm de}(a)$, this evolution is given by:
\begin{equation}
\rho_{\rm de}(a) = \rho_{{\rm de},0}
\exp\left[-3\int_1^a \left(1+w_{\rm de}(\tilde a)\right)d\ln \tilde a\right] \, ,
\label{eq:rho_de_background}
\end{equation}
where $a=a(t)$ is the scale factor,  $\rho_{{\rm de},0}$ the present-day dark energy density. For a spatially flat Friedmann–Lemaître–Robertson–Walker
(FLRW) background, the Hubble parameter is given by
\begin{equation}
\frac{H^2(a)}{H_0^2} =
\Omega_{{\rm r},0}a^{-4}+\Omega_{{\rm m},0}a^{-3}
+\Omega_{{\rm de},0}\frac{\rho_{\rm de}(a)}{\rho_{{\rm de},0}} \, ,
\label{eq:friedmann_background}
\end{equation}
where $H_0$ is the present-day Hubble rate,  $\Omega_{{\rm r},0}$, $\Omega_{{\rm m},0}$, and $\Omega_{{\rm de},0}$  denote the present-day density parameters of  radiation, matter, and dark energy, respectively.

\subsection{Early Dark Energy}
\label{EDE_model}

The primary parametrization considered in this work belongs to the class of unified EDE models, introduced by Doran and Robbers \cite{2006JCAP...06..026D} (see also \cite{LINDER200616}). In this approach, the fluid tracks the EoS of the dominant matter component of the Universe at early times, behaving like radiation ($w_{\rm ede} \sim 1/3$) during radiation domination and like matter ($w_{\rm ede} \sim 0$) during matter domination epoch, while maintaining a non-negligible fractional density $\Omega_e^{\rm EDE}$ before transitioning to drive late-time cosmic acceleration. The effective EoS, $w_{\rm ede}(a)$, is modeled by
\begin{equation}
w_{\rm ede} (a) = - \frac{\dfrac{d \ln{\Omega_{\rm ede }(a)}}{d \ln{a}}}{3 \left[ 1 - \Omega_{\rm ede }(a) \right]}  + \frac{a_{\rm eq}}{3 \left(a + a_{\rm eq} \right)} \, ,
\label{eq:eos_parameter}
\end{equation}
where the EDE fractional density parameter evolves as
\begin{equation}
    \begin{split}
        \Omega_{\rm ede} (a) = & \ \dfrac{\Omega_{\rm de,0} - \Omega_e^{\rm EDE} \left( 1- a^{- 3 w_0}\right)}{\Omega_{\rm de,0} + \Omega_{\rm m,0} \ a^{3 w_0}} \\
        & + \Omega_e^{\rm EDE} (1-a^{-3 w_0}) \, .
    \end{split}
    \label{eq:density_param}
\end{equation}
Here, $a_{\rm eq}$ is the scale factor at matter-radiation equality, while $w_0$ and $\Omega_e^{\rm EDE}$ are two additional free parameters.

The parameter $\Omega_e^{\rm EDE}$ sets the asymptotic early-time dark energy fraction, such that $\Omega_{\rm ede}(a) \to \Omega_e^{\rm EDE}$ as $a \to 0$ (or, equivalently, $z \to \infty$), thereby determining the fractional contribution of EDE to the total energy density well before recombination.

\begin{figure}
\centering
\includegraphics[width=1.0\linewidth]{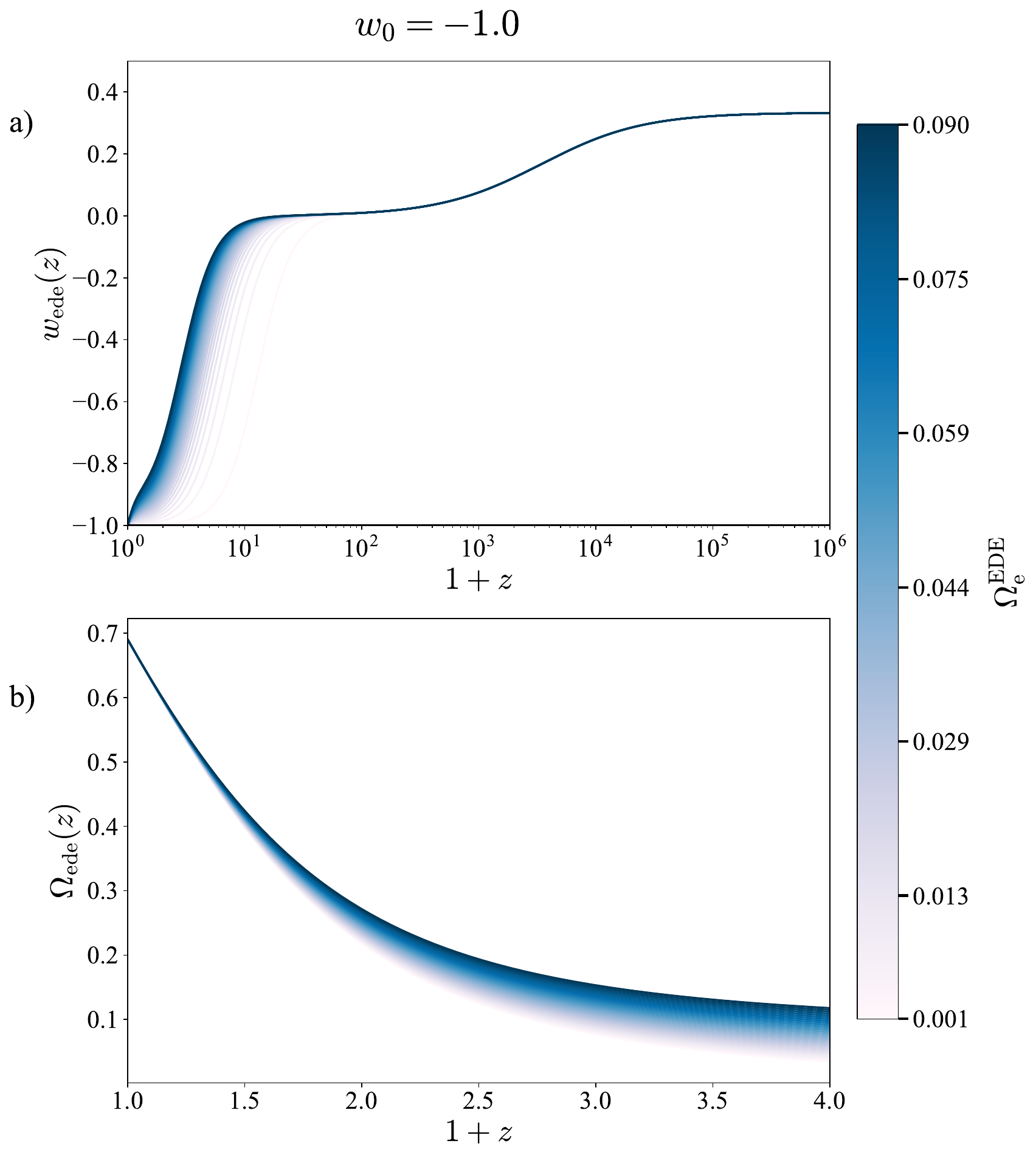}
\caption{Cosmological background dynamics of the EDE fluid assuming $w_0 = -1$, for varying early-time energy density $\Omega_{\rm e}^{\rm EDE} \in [0.001, 0.09]$. \textbf{a)} We plot the evolution of $w_{\rm de}(z)$. Darker shades correspond to higher $\Omega_{\rm e}^{\rm EDE}$ values. The solid black line marks the standard $\Lambda$CDM limit at present time. \textbf{b)} Redshift evolution of the dimensionless EDE density. }
\label{fig:background_evolution}
\end{figure}

The effect of $\Omega_{\rm e}^{\rm EDE}$ on the cosmic evolution for a fixed $w_0 = -1.0$ is illustrated in Fig.~\ref{fig:background_evolution}. As required to match late-time observations, the upper panel shows that the model recovers a cosmological constant behavior today, with $w_{\rm ede} \to -1$ as $z \to 0$. Moving backward in time, however, the fluid undergoes the transition described above, allowing $w_{\rm ede } > -1$ at higher redshifts. Consequently, unlike a cosmological constant, its energy density evolves with redshift, allowing EDE to constitute a non-negligible fraction of the total energy density at early times. The magnitude of this contribution is controlled by $\Omega_{\rm e}^{\rm EDE}$, as illustrated in the bottom panel.

\subsection{Late Dark Energy}
\label{CPL_model}

As a complementary scenario, we consider the well-known CPL parametrization \cite{CHEVALLIER_2001, Linder},
\begin{equation}
w_{\rm CPL}(a) = w_0 + w_a (1 - a)\, ,
\label{eq:cpl_param}
\end{equation}
where $w_0$ is the present EoS value and $w_a$ controls its  evolution. 
CPL belongs to a broad family of late-time dark energy parametrizations, with $w_{\rm de}<-1/3$ ensuring that the dark energy density grows more slowly than the matter density toward the past and thus becomes negligible at early times. It provides a simple description of dynamical departures from $w=-1$, while recovering $\Lambda$CDM for $(w_0,w_a)=(-1,0)$. Despite their different functional forms, several late-time parametrizations yield similar trends \cite{DESI:2025fii}, making CPL a simple and representative choice for our analysis. In the early-time limit, the EoS approaches $w_{\rm CPL}(a\to0)=w_0+w_a$.

\section{Dark Energy perturbations}
\label{sec:DE_perturbations}

To account for linear perturbations, we consider scalar perturbations around a spatially flat FLRW background in the Newtonian gauge, with line element:
\begin{equation}
    ds^2 = -(1+2\Psi)dt^2 + a^2(t)(1-2\Phi)d\mathbf{x}^2 \, ,
    \label{eq:newtonian_metric}
\end{equation}
where $\Psi$ and $\Phi$ denote the gravitational metric potentials, and $\mathbf{x}$ represents the spatial comoving coordinates. At linear order, the hydrodynamic evolution of dark energy fluctuations is dictated directly by energy-momentum conservation, holding independently of the specific functional form assigned to $w_{\rm de}(z)$. For a non-interacting dark energy component, this leads to the standard continuity and Euler equations \cite{1995ApJ...455....7M}:
\begin{align}
& \delta\rho_{\mathrm{de}}' + 3\bigl(\delta\rho_{\mathrm{de}} + \delta P_{\mathrm{de}}\bigr)
= -\bigl(\rho_{\mathrm{de}} + P_{\mathrm{de}}\bigr)
\left(3\Phi' + \frac{\theta_{\mathrm{de}}}{aH}\right), \\
&\bigl(\rho_{\mathrm{de}} + P_{\mathrm{de}}\bigr)\theta_{\mathrm{de}}'
= \frac{k^2}{aH}
\left[\Psi\bigl(\rho_{\mathrm{de}} + P_{\mathrm{de}}\bigr) + \delta P_{\mathrm{de}}\right] \nonumber \\
& - \bigl(\rho_{\mathrm{de}}' + P_{\mathrm{de}}'\bigr)\theta_{\mathrm{de}}
- 4\bigl(\rho_{\mathrm{de}} + P_{\mathrm{de}}\bigr)\theta_{\mathrm{de}} ,
\end{align}
where primes denote derivatives with respect to $\ln a$, and $\theta_{\mathrm{de}} \equiv \nabla \cdot \vec{v}_{\mathrm{de}}$ is the velocity divergence.
For this component, the pressure perturbation expands as
\begin{equation}
\delta P_{\mathrm{de}} = c_s^2\,\delta\rho _{\mathrm{de}}+ 3aH(\rho_{\mathrm{de}} + P_{\mathrm{de}})(c_s^2 - c_a^2)\frac{\theta_{\mathrm{de}}}{k^2},
\end{equation}
in terms of the adiabatic sound speed $c_a^2 = P'_{\rm de}/\rho'_{\rm de}$ and the rest-frame sound speed $c_s^2 = \delta P_{\rm de}^{\rm rest}/\delta \rho_{\rm de}^{\rm rest}$. Setting $c_s^2 = 1$, we numerically evolve the dark energy perturbations using the Parametrized Post-Friedmann (PPF) scheme \cite{Hu:2007pj, Hu:2008zd, Fang:2008sn}, which provides a regular description across the phantom divide and consistently connects the perturbation evolution on large and small scales. 

\begin{figure*}[!htbp]
    \centering    \includegraphics[width=0.95\linewidth]{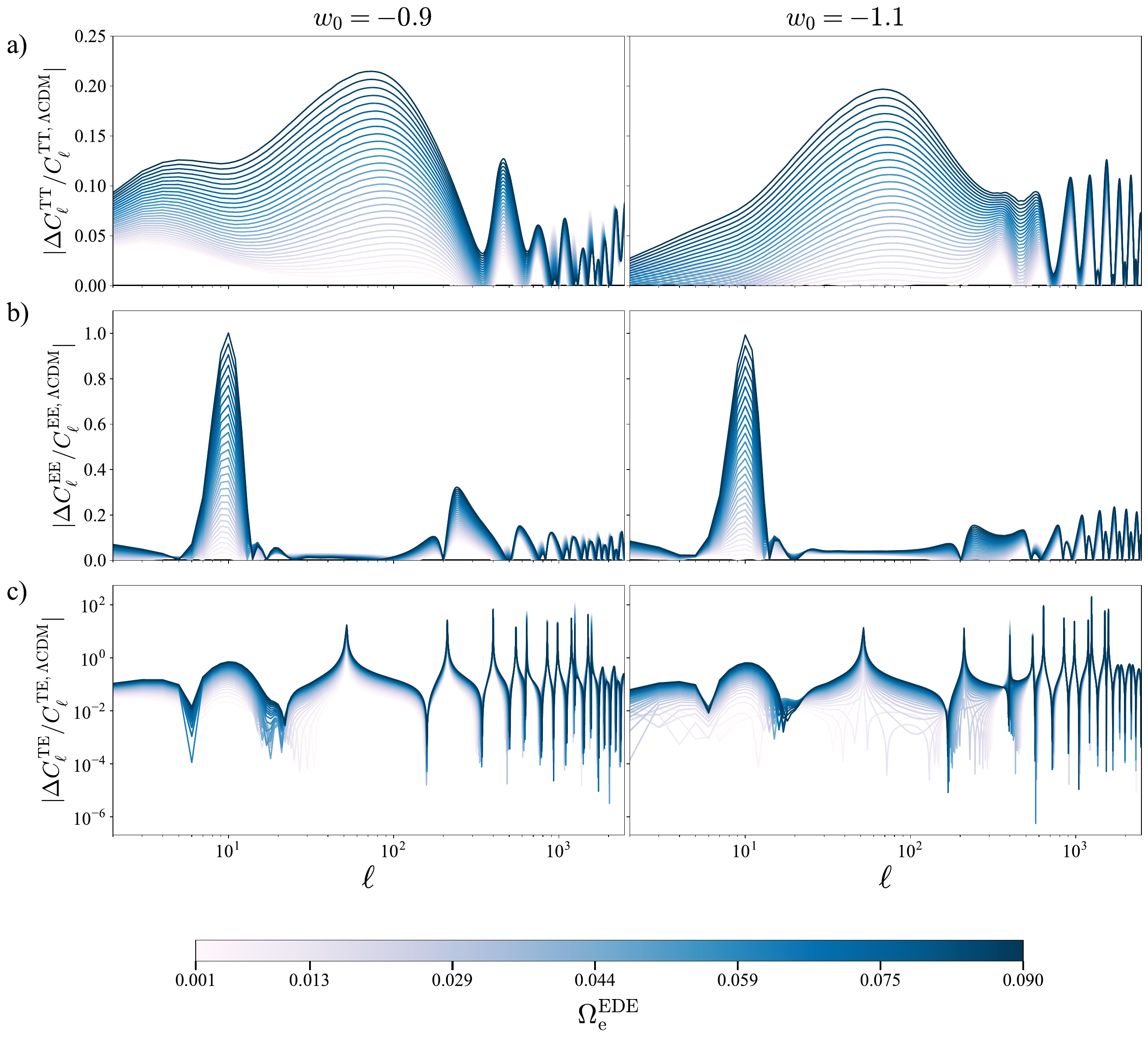}
    \caption{Relative differences in the CMB power spectra,  $\Delta C^{\rm XY}_\ell =  C^{\rm XY, \ EDE}_\ell - C^{\rm XY, \ \Lambda CDM}_\ell$, showing the deviation of the EDE model with respect to $\Lambda$CDM, a) the temperature auto-correlation, b) polarization auto-correlation, and c) temperature-polarization cross-correlation. The left column corresponds to a fixed $w_0 = -0.9$, while the right column corresponds to $w_0 = -1.1$. The color gradient indicates the fraction of EDE $\Omega^{\rm EDE}_{\rm e}$, ranging from $0.001$ to $0.09$. The plots demonstrate how varying $\Omega_{\text{EDE}}$ induces scale-dependent deviations in the spectra, particularly affecting the acoustic peak structure.}    
    \label{fig:cmb_spectra_ede}
\end{figure*}

The resulting evolution allows us to track the impact of dark energy perturbations on the gravitational potentials $\Phi$ and $\Psi$. These metric perturbations, along with the modified background expansion history, induce specific signatures in the CMB anisotropies, particularly in the EDE case. We illustrate these effects in Fig.~\ref{fig:cmb_spectra_ede}, which displays the relative deviations in the temperature (TT), polarization (EE), and cross-correlation (TE) power spectra with respect to the standard $\Lambda$CDM model. As seen in the figure, for both values of $w_0$ considered, increasing the EDE fraction, $\Omega^{\rm EDE}_{\rm e}$, progressively enhances the deviations from the $\Lambda$CDM spectra, with larger EDE fractions producing more pronounced modifications to the acoustic peak amplitudes, providing the distinct spectral fingerprints necessary to constrain the model with CMB data. As reference, the impact on the matter power spectrum is shown in Fig.~\ref{fig:matter_pk} of Appendix~\ref{app:A}.

\section{Methodology and Dataset}
\label{sec:Methods&Data}

In this section, we present the details of our analysis.
The evolution of background and perturbation equations for both dark energy parametrizations was computed using the Boltzmann code \texttt{CLASS} \cite{2011JCAP...07..034B, 2011arXiv1104.2932L}.  For CPL  we use the standard implementation, while the EDE model was incorporated through a modified version of the code.\footnote{\url{https://github.com/mazapataa/EDE}}

\begin{table}[!htbp]
\caption{Priors of cosmological parameters for inference analysis, used in the MCMC runs that include CMB data (Sec.~\ref{CMB}); the background-only runs of Sec.~\ref{pure_background} use the narrower ranges listed in Tab.~\ref{tab:table_priors_bg}}. Each parameter has a flat top-hat distribution.
\setlength{\tabcolsep}{10pt} 
\renewcommand{\arraystretch}{1.5}
\begin{tabular}{@{}l@{\hspace{3em}}cc@{}} 
\toprule
\toprule
Parameter & Prior Range & Distribution \\ 
\midrule
$\omega_{\rm b} = \Omega_{\rm b} h^2$       & $[0.005, 0.1]$     & \tophattails \\
$\omega_{\rm cdm} = \Omega_{\rm cdm} h^2$   & $[0.001, 0.99]$  & \tophattails \\
$100 \, \theta_{\rm s}$                     & $[0.5, 10]$      & \tophattails \\
$\tau_{\rm reio}$                           & $[0.01, 0.8]$    & \tophattails \\
$\ln \left(10^{10} A_{\rm s} \right)$       & $[1.61, 3.91]$   & \tophattails \\
$\sum m_{\nu}$ [eV]                               & $[0, 2]$     & \tophattails \\
$n_s$                                       & $[0.8, 1.2]$     & \tophattails \\
$w_0$                                       & $[-3, 1]$        & \tophattails \\
$\Omega_{\rm e}^{\rm EDE}$                  & $[0.0, 0.5]$     & \tophattails \\
$w_{\rm a}$                                & {$[-3,2]$}   & \tophattails \\
\bottomrule
\bottomrule
\end{tabular}
\label{tab:table_priors}
\end{table}

\subsection{Bayesian Inference}
\label{Bayesian_inference}

To determine the posterior distributions of the cosmological parameters for the EDE and CPL models, we conduct a Bayesian analysis combining different observational datasets. 
We assigned flat priors on the cosmological and model parameters, as specified in Tab.~\ref{tab:table_priors}. For dataset combinations including CMB observations, the inference is performed using the \texttt{Cobaya}\footnote{\url{https://github.com/CobayaSampler/cobaya}} \cite{Torrado:2020dgo} framework. {In every run the neutrino sector is treated identically across the three models: we assume a single massive eigenstate ($N_{\rm ncdm}=1$) together with $N_{\rm ur}=2.0328$ massless species, giving $N_{\rm eff}=3.046$, and we vary $\sum m_\nu$ freely. Keeping this sector fixed between $\Lambda$CDM and its extensions is essential for the model comparison to be meaningful.}

For inferences involving purely geometric distance measurements (BAO and SNe Ia), parameter sampling and evidence estimation are instead performed using the MLFriends algorithm~\cite{2016S&C....26..383B,2019PASP..131j8005B} implemented in the nested sampling code: \texttt{UltraNest}\footnote{\url{https://github.com/JohannesBuchner/UltraNest}}~\cite{2021JOSS....6.3001B}. This algorithm directly computes the Bayesian evidence (marginal likelihood), defined as \cite{Trotta:2008qt}: 
\begin{equation}
    \mathcal{Z} \equiv P(\mathcal{D}|\mathcal{M}) = \int \mathcal{L}(\boldsymbol{\theta})\,\pi(\boldsymbol{\theta})\,\mathrm{d}\boldsymbol{\theta}\,,
\end{equation}
where $\mathcal{L}(\boldsymbol{\theta}) \equiv P(\mathcal{D}|\boldsymbol{\theta}, \mathcal{M})$ is the likelihood of the data $\mathcal{D}$ given the parameter vector $\boldsymbol{\theta}$ under model $\mathcal{M}$, and $\pi(\boldsymbol{\theta}) \equiv P(\boldsymbol{\theta}|\mathcal{M})$ represents the prior probability distribution. While $\mathcal{Z}$ acts as a normalization factor in parameter estimation, it provides a key quantity for Bayesian model comparison \cite{Liddle:2007fy}.

To evaluate the relative preference that data $\mathcal{D}$ lend to a candidate model $\mathcal{M}_1$ over a reference hypothesis $\mathcal{M}_2$, we compute the Bayes factor, defined as the ratio of their marginal likelihoods:
\begin{equation}
 B_{12} = \frac{\mathcal{Z}_{\mathcal{M}_1}}{\mathcal{Z}_{\mathcal{M}_2}} = \frac{\mathcal{P} ( \mathcal{D} \vert \mathcal{M}_1 )}{ \mathcal{P} ( \mathcal{D} \vert \mathcal{M}_2 )}. 
 \label{eq:bayes_factor} 
\end{equation}
In practice, we report the difference in log-evidence between models, $\Delta\ln\mathcal{Z}=\ln B_{12}$, and quantify the strength of the statistical preference using Jeffreys' scale from Tab.~\ref{tab:jeffreys_bellido}~\citep{jeffreys1998theory,Nesseris:2012cq}, which provides a standardized benchmark to categorize evidence levels from weak to very strong. 

\begin{table}[!htbp]
\centering
\caption{Model comparison metrics between $\Lambda$CDM and dynamical dark energy models. Here, $\Delta \ln \mathcal{Z}$ is interpreted via Jeffreys' scale for Bayesian model comparison as defined in \cite{Nesseris:2012cq}.}
\label{tab:jeffreys_bellido}
\begin{tabular*}{0.47\textwidth}{@{\extracolsep{\fill}}ccc}
\toprule 
\textbf{Bayes Factor ($B_{ij}$)} & \textbf{$|\Delta \ln \mathcal{Z}|$} & \textbf{Evidence} \\ 
\midrule
$1 \le B_{ij} < 3$     & $0 \le |\Delta \ln \mathcal{Z}| < 1.1$ & Weak \\
$3 \le B_{ij} < 20$    & $1.1 \le |\Delta \ln \mathcal{Z}| < 3.0$ & Definite \\
$20 \le B_{ij} < 150$  & $3.0 \le |\Delta \ln \mathcal{Z}| < 5.0$ & Strong \\
$B_{ij} \ge 150$       & $|\Delta \ln \mathcal{Z}| \ge 5.0$      & Very Strong \\
\bottomrule
\end{tabular*}
\end{table}

\subsection{Evidence Estimation from MCMC Chains via \texttt{harmonic}}

For dataset combinations including CMB data, where high-dimensional parameter spaces make nested sampling computationally unfeasible in our pipeline, we perform parameter inference using Markov Chain Monte Carlo (MCMC) sampler \cite{Lewis:2002ah} within \texttt{Cobaya}. To evaluate the Bayesian evidence $\mathcal{Z}$ from these chains, we employ the learned harmonic mean estimator implemented in \texttt{harmonic}\footnote{\url{https://github.com/astro-informatics/harmonic}} \cite{harmonic,polanska2024learned}, a post-processing approach increasingly adopted in cosmological model selection \cite[see e.g.,][]{Piras:2024dml,Carrion:2024jur,Lin:2025xbw}.

Unlike nested sampling, which computes $\mathcal{Z}$ directly during runtime, \texttt{harmonic} reconstructs the evidence purely from posterior samples without requiring further likelihood evaluations. It bypasses the severe variance instability of the classic harmonic mean estimator \cite{Newton1994},
\begin{equation}
\frac{1}{\hat{\mathcal{Z}}} = \frac{1}{N}\sum_{i=1}^{N} \frac{1}{\mathcal{L}(\boldsymbol{\theta}_i)}\,, \qquad \boldsymbol{\theta}_i \sim P(\boldsymbol{\theta}|\mathcal{D})\,,
\end{equation}
which is typically dominated by rare samples in low-likelihood regions. To resolve this pathology, \texttt{harmonic} introduces a learned target distribution $\varphi(\boldsymbol{\theta})$, trained directly on posterior samples via normalizing flows such that its density remains safely confined within the bulk of the posterior. The evidence is then reliably estimated via
\begin{equation}
\frac{1}{\mathcal{Z}_{\rm H}} \approx \frac{1}{N}\sum_{i=1}^{N} \frac{\varphi(\boldsymbol{\theta}_i)}{\mathcal{L}(\boldsymbol{\theta}_i)\,\pi(\boldsymbol{\theta}_i)}\,,
\label{eq:harmonic_estimator}
\end{equation}
which converges to an unbiased, finite-variance estimate of $1/\mathcal{Z}$ provided $\varphi(\boldsymbol{\theta})$ satisfies the required normalization and support constraints. Because it requires only posterior samples, \texttt{harmonic} is much cheaper to run than nested sampling once a chain already exists, making it the practical choice for the CMB-included analyses of Subsec.~\ref{CMB}. This comes with a caveat worth stating plainly: the uncertainties quoted for $\ln\mathcal{Z}$ by \texttt{UltraNest} and by \texttt{harmonic}, though written with the same notation here, are not the same statistical quantity. \texttt{UltraNest} uncertainty comes from the stochastic shrinkage of the prior volume during the run itself, and reflects the precision of one self-contained estimate of $\mathcal{Z}$. While \texttt{harmonic} uncertainty is instead the Monte Carlo variance of the importance-sampling estimator in Eq.~\ref{eq:harmonic_estimator}, computed over a fixed, already existing set of posterior samples. A small \texttt{harmonic} error bar therefore does not carry the same weight as an equally small nested sampling one. This is why Sec.~\ref{pure_background} cross-validates the two methods directly, wherever both are feasible, finding that the two agree on $\ln B$ to better than $0.25$ across all model and dataset combinations where both are available; it is this agreement, not the quoted errors on their own, that justifies using \texttt{harmonic} for the CMB-included evidences of Subsec.~\ref{CMB}, where no such direct check is possible.

Our analysis constrains cosmological parameters using the following datasets: BAO measurements from DESI, the Pantheon+ Type Ia supernova catalog, and the full CMB data from \textit{Planck} 2018.  

\begin{itemize}
    \item[$\star$] \textit{Baryon Acoustic Oscillations} \textbf{(DESI DR2):} We use the consensus BAO measurements from DESI Data Release 2 \cite{DESI:2025zgx}. These measurements, from galaxies and quasars, are expressed in terms of the transverse comoving distance $D_M(z)/r_d$ and the Hubble distance $D_H(z)/r_d$, both normalized by the sound horizon at the baryon drag epoch, $r_d$. These measurements cover the redshift range $0.295 \leq z \leq 2.33$. The full covariance matrix, which includes both statistical and systematic uncertainties, is incorporated within \texttt{Cobaya}. We refer to this dataset as \textbf{DESI}. \\
    
    \item[$\star$] \textit{Type Ia Supernovae} \textbf{(Pantheon+):} We utilize the distance modulus measurements from the Pantheon+ sample \cite{Scolnic:2021amr}, which provide the most comprehensive compilation of Type Ia supernova observations, featuring 1701 high-quality light curves from 1550 distinct supernovae. 
    This dataset spans an extensive redshift range from $0.01 < z < 2.26$. We designate this dataset as \textbf{PP}. \\

    \item[$\star$] \textit{Cosmic Microwave Background} (\textit{Planck}): We employ the final \textit{Planck} 2018 legacy data \cite{Planck:2019nip, Planck:2018vyg} release through the following native \texttt{Cobaya} likelihoods: the low-$\ell$ of the temperature power spectrum TT over the multipole range $2 \leq \ell \leq 29$, the low-$\ell$ of the polarization power spectrum EE from $2 \leq \ell \leq 29$, the high-$\ell$ of the TT, TE, and EE power spectra spanning $30 \leq \ell \leq 2508$, and the CMB lensing \cite{2020A&A...641A...8P} reconstruction derived from the full temperature and polarization data. 
    We label this data compilation as \textit{Planck}.
    
\end{itemize}

\section{Results and Discussion}
\label{Results}

\begin{table*}[!htbp]
\centering
\begin{threeparttable}
\caption{Mean values of cosmological parameters for the $\Lambda$CDM, EDE, and CPL models from the \textbf{DESI} and \textbf{PP} datasets. Quoted uncertainties are at $68\%$ confidence level, while upper/lower bounds are at $95\%$. Evidence values are obtained from nested sampling (background-only fits) or post-processed via \texttt{harmonic} (CMB-included fits); the two agree on the Bayes factor to within $0.24$ in $\ln B$ across all four model, dataset combinations (largest discrepancy for EDE, \textbf{DESI}; smallest, $0.07$, for CPL, \textbf{DESI}), validating \texttt{harmonic} against \texttt{UltraNest}. Dashes indicate
parameters not applicable to the corresponding model. An ellipsis ($\cdots$) indicates a parameter that remains unconstrained by the corresponding dataset.}

\label{tab:cosmo_parameter_constraints}
\renewcommand{\arraystretch}{1.35}
\setlength{\tabcolsep}{4pt}
\footnotesize
\begin{tabular}{l cc cc cc}
\toprule
& \multicolumn{2}{c}{\textbf{$\Lambda$CDM}}
& \multicolumn{2}{c}{\textbf{EDE}}
& \multicolumn{2}{c}{\textbf{CPL}} \\
\cmidrule(lr){2-3} \cmidrule(lr){4-5} \cmidrule(lr){6-7}
 & \textbf{DESI} & \textbf{DESI+PP} & \textbf{DESI} & \textbf{DESI+PP} & \textbf{DESI} & \textbf{DESI+PP} \\
\midrule
\multicolumn{7}{l}{\textbf{Baseline parameters}} \\
\addlinespace[2pt]
$h$
& $0.687 \pm 0.0042$ & $0.687 \pm 0.004$
& $0.703 \pm 0.019$ & $0.692 \pm 0.015$
& $0.663^{+0.011}_{-0.019}$ & $0.677 \pm 0.009$ \\
$\Omega_{\rm m}$
& $0.298^{+0.007}_{-0.009}$ & $0.304 \pm 0.008$
& $0.284 \pm 0.013$ & $0.289 \pm 0.012$
& $0.328^{+0.021}_{-0.012}$ & $0.311 \pm 0.009$ \\
\addlinespace[3pt]
\midrule
\multicolumn{7}{l}{\textbf{Dark energy parameters}} \\
\addlinespace[2pt]
$w_0$
& -- & --
& $-0.991^{+0.060}_{-0.049}$ & $-0.953 \pm 0.030$
& ${-0.707^{+0.207}_{-0.052}}$ & $-0.876 \pm 0.058$ \\
$w_a$
& -- & --
& -- & --
& $-0.89^{+0.36}_{-0.61}$ & $-0.39^{+0.31}_{-0.26}$ \\
$\Omega_e^{\rm EDE}$
& -- & --
& $\cdots$ & $\cdots$
& -- & -- \\
\addlinespace[3pt]
\midrule
\multicolumn{7}{l}{\textbf{Statistics}} \\
\addlinespace[2pt]
$\chi^2_{\rm mean}$
& $12.05$ & $1418.09$
& $12.12$ & $1414.98$
& $9.86$ & $1414.67$ \\
$-\ln\mathcal{Z}$
& $8.93 \pm 0.09$ & $711.87 \pm 0.12$
& $10.90 \pm 0.12$ & $713.35 \pm 0.13$
& $10.81 \pm 0.17$ & $714.23 \pm 0.21$ \\
$-\ln\mathcal{Z}_{\rm H}$
& $9.01 \pm 0.05$ & $712.03 \pm 0.02$
& $11.22 \pm 0.07$ & $713.34 \pm 0.05$
& $10.82 \pm 0.03$ & $714.28 \pm 0.02$ \\
$\ln B_{\Lambda{\rm CDM},i}$
& -- & --
& $1.97 \pm 0.15$ & $1.48 \pm 0.18$
& $1.88 \pm 0.19$ & $2.36 \pm 0.24$ \\
\bottomrule
\end{tabular}
\end{threeparttable}
\end{table*}

\subsection{Background analysis}
\label{pure_background}

Before turning to the full CMB-included analysis, we evaluate the constraints obtained from purely geometric probes: \textbf{DESI} BAO alone and in combination with \textbf{PP}. The parameter constraints, goodness-of-fit, and Bayesian evidences for $\Lambda\text{CDM}$, EDE, and CPL are summarized in Tab.~\ref{tab:cosmo_parameter_constraints}. This background-only step serves two key purposes: first, it isolates how low-redshift geometric probes respond to early- and late-time dark energy dynamics without the tight anchor of CMB anisotropies; second, since evidences on background likelihoods can be computed directly via \texttt{UltraNest} nested sampling, it provides a rigorous benchmark to cross-validate the \texttt{harmonic} post-processing estimator before deploying it on full-likelihood MCMCs in Sec.~\ref{CMB}.

A prominent physical feature of both extended models at the background level is the additional freedom they introduce in the late-time expansion history; for \textbf{DESI} data alone, EDE shifts the preferred expansion rate toward larger values, $h=0.703\pm0.019$, accompanied by a lower matter fraction, $\Omega_{\rm m}=0.284\pm0.013$, compared with $h=0.687\pm0.004$ and $\Omega_{\rm m}=0.298^{+0.007}_{-0.009}$ in $\Lambda$CDM.  
Interestingly, this shift occurs even though the present-day EoS remains remarkably close to that of a cosmological constant.
CPL instead moves in the opposite direction, favoring a lower $h=0.663^{+0.011}_{-0.019}$ and a larger $\Omega_{\rm m}=0.328^{+0.021}_{-0.012}$. Once \textbf{PP} is included, these differences are substantially reduced: the inferred values become $h=0.692\pm0.015$ and $0.677\pm0.009$ for EDE and CPL, respectively, while their matter fractions move closer to the $\Lambda$CDM value.

Regarding the parameters controlling the dark energy dynamics, the analysis shows that in the EDE model late-time data alone do not constrain the early dark energy fraction, $\Omega_e^{\rm EDE}$. In contrast, $w_0$ remains close to $-1$ for both \textbf{DESI} and \textbf{DESI}+\textbf{PP}, with the cosmological constant value lying only about $0.2\sigma$ and $1.6\sigma$ away from the corresponding marginalized constraints, respectively.
For CPL, \textbf{DESI} alone leaves the dark energy dynamics poorly constrained, while the inclusion of \textbf{PP} yields $w_0=-0.876\pm0.058$ and $w_a=-0.39^{+0.31}_{-0.26}$, corresponding to deviations from $w_0=-1$ and $w_a=0$ of approximately $2.1\sigma$ and $1.3\sigma$, respectively.

Overall, the model selection results show that the modest reduction in $\chi^2$ achieved by the extended models does not translate into a Bayesian preference over $\Lambda$CDM. Despite slightly lower $\chi^2$ values, particularly after the inclusion of \textbf{PP}, the direct nested-sampling Bayes factors consistently favor $\Lambda$CDM across all background dataset combinations, with $\ln B_{\Lambda{\rm CDM},i}$ ranging from approximately $1.5$ to $2.4$. Thus, improvements of up to $\Delta\chi^2_{\rm mean}\simeq-3.4$ are not sufficient to overcome the Bayesian Occam penalty associated with the additional parameter freedom. This background level preference for $\Lambda$CDM agrees with recent independent nested sampling analyses~\cite{Ong:2025utx}, indicating that apparent preferences for dynamical dark energy can be weakened once the full prior volume is taken into account.

Finally, the direct nested sampling evidences and the post-processed harmonic estimates agree on $\ln B_{\Lambda{\rm CDM},i}$ to better than $0.24$ across the four background combinations where both are available; a small systematic offset in $\ln\mathcal{Z}$ itself is common to all models and cancels in the ratio. This justifies applying harmonic to the CMB-included likelihoods of Sec.~\ref{CMB}, where direct nested sampling is computationally unfeasible.

\begin{table*}[!htbp]
\begin{threeparttable}
\caption{Marginalized mean values (or upper limits) for cosmological parameters
in the EDE and CPL extensions to $\Lambda$CDM, obtained from \textit{Planck} alone and in
combination with \textbf{DESI} and \textbf{PP}.
Two-sided uncertainties are $68\%$ confidence level, rounded to one
significant figure, with the corresponding central value rounded to the same
decimal place; upper limits are quoted at $95\%$ confidence level to two significant figures. Parameters whose natural magnitude is $\lesssim 0.1$ are rescaled by the indicated power of ten for readability. Dashes indicate
parameters not applicable to the corresponding model; an ellipsis ($\cdots$) indicates a parameter that remains unconstrained by the corresponding dataset. Evidences $-\ln\mathcal{Z}_{\rm H}$
are the mean of five independent realisations of the \texttt{harmonic} estimator
(fixed seeds $0$--$4$), and the quoted uncertainty is the scatter across realisations.}
\label{tab:results_ede_cpl_comparison}
\centering
\renewcommand{\arraystretch}{1.45}
\setlength{\tabcolsep}{3.0pt}
\begin{tabular*}{1.0\textwidth}{@{\extracolsep{\fill}}l ccc ccc}
\toprule
& \multicolumn{3}{c}{\textbf{EDE}} & \multicolumn{3}{c}{\textbf{CPL}} \\
\cmidrule(lr){2-4} \cmidrule(lr){5-7}
\multicolumn{1}{c}{} & \textit{Planck} & \textbf{+DESI} & \textbf{+DESI+PP} & \textit{Planck} & \textbf{+DESI} & \textbf{+DESI+PP} \\
\midrule
\multicolumn{7}{l}{\textbf{Baseline Parameters}} \\
$h$ & $0.63^{+0.03}_{-0.02}$ & $0.679^{+0.006}_{-0.005}$ & $0.676\pm 0.005$ & $0.9\pm 0.2$ & $0.64\pm 0.02$ & $0.676\pm 0.006$ \\
$100\,\omega_b$ & $2.23\pm 0.02$ & $2.26\pm 0.01$ & $2.26\pm 0.01$ & $2.24\pm 0.02$ & $2.24\pm 0.01$ & $2.24\pm 0.01$ \\
$100\,\omega_{\rm cdm}$ & $12.2\pm 0.1$ & $11.78\pm 0.08$ & $11.77\pm 0.08$ & $12.0\pm 0.1$ & $11.95\pm 0.09$ & $11.90\pm 0.09$ \\
$\log \left(10^{10} A_s \right)$ & $3.05\pm 0.02$ & $3.06\pm 0.02$ & $3.06^{+0.01}_{-0.02}$ & $3.04\pm 0.02$ & $3.04\pm 0.02$ & $3.04\pm 0.02$ \\
$100\,n_s$ & $96.2\pm 0.5$ & $97.1\pm 0.3$ & $97.1\pm 0.3$ & $96.5\pm 0.5$ & $96.6\pm 0.4$ & $96.7\pm 0.4$ \\
$\sum m_{\nu}$ [eV] & $< 0.34$ & $< 0.050$ & $< 0.049$ & $< 0.40$ & $< 0.18$ & $< 0.13$ \\
$100\,\tau_{\rm reio}$ & $5.5\pm 0.8$ & $6.4\pm 0.8$ & $6.3^{+0.7}_{-0.8}$ & $5.3\pm 0.7$ & $5.4\pm 0.8$ & $5.5\pm 0.8$ \\
\midrule
\multicolumn{7}{l}{\textbf{Dark Energy Parameters}} \\
$w_0$ & $-0.89^{+0.04}_{-0.09}$ & $-0.97^{+0.01}_{-0.03}$ & $-0.96\pm 0.02$ & $-1.7^{+0.6}_{-0.8}$ & $-0.4\pm 0.2$ & $-0.85\pm 0.06$ \\
$w_a$ & -- & -- & -- & $\cdots$ & $-1.8\pm 0.6$ & $-0.6\pm 0.2$ \\
$10^{3}\,\Omega_{\rm e}^{\rm EDE}$ & $< 5.3$ & $2.4^{+0.9}_{-1.0}$ & $2.0^{+0.6}_{-1.0}$ & -- & -- & -- \\
\midrule
\multicolumn{7}{l}{\textbf{Statistics values}} \\
$\chi^2_{\rm mean}$ & $2793.91$ & $2809.70$ & $4213.13$ & $2785.51$ & $2796.04$ & $4201.83$ \\
$-\ln \mathcal{Z}_{\rm H}$ & $1461.42 \pm 0.04$ & $1473.75 \pm 0.08$ & $2175.39 \pm 0.04$ & $1450.71 \pm 0.14$ & $1460.23 \pm 0.06$ & $2165.12 \pm 0.08$ \\
$\ln B_{\Lambda{\rm CDM},i}$
& $9.90 \pm 0.09$ & $13.97 \pm 0.11$ & $12.61 \pm 0.17$
& $-0.81 \pm 0.16$ & $0.45 \pm 0.10$ & $2.34 \pm 0.18$ \\
\bottomrule
\end{tabular*}
\end{threeparttable}
\end{table*}

\subsection{Including CMB}
\label{CMB}

Turning to the joint analysis, we evaluate how incorporating \textit{Planck} CMB anisotropy data alongside \textbf{DESI} BAO and \textbf{PP} improves the constraints and evidence for both models. Likewise parameter constraints, goodness-of-fit metrics, and Bayes factors are summarized in Tab.~\ref{tab:results_ede_cpl_comparison}.

\begin{figure}[!htbp]
\centering
\includegraphics[width=0.48\textwidth]{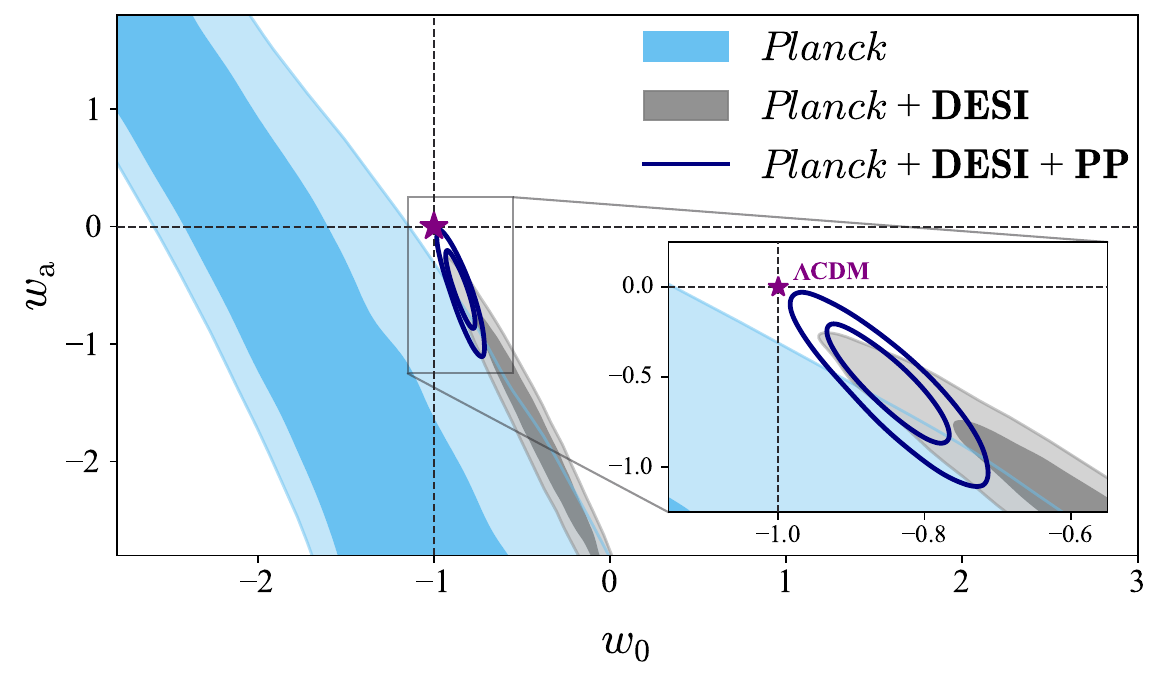}
\caption{Marginalized $68\%$ and $95\%$ C.L. constraints on the CPL dark energy parameter space ($w_0, w_a$). The broad geometric degeneracy allowed by \textit{Planck} CMB data alone (light blue) is significantly broken upon combining with low-redshift measurements from \textbf{DESI} (grey) and the full dataset including \textbf{PP} (navy contours). The standard $\Lambda\text{CDM}$ cosmology ($w_0 = -1, w_a = 0$) is indicated by the purple star. The inset panel provides a zoomed-in view of the region surrounding $\Lambda\text{CDM}$, clearly illustrating the displacement of the joint posterior distributions from the standard cosmological model.}
\label{fig:cpl_contours_2D}
\end{figure}

Fig.~\ref{fig:cpl_contours_2D} highlights the 2D posterior distributions for the CPL parametrization. As expected, \textit{Planck} data alone (light blue contour, $w_0=-1.7^{+0.6}_{-0.8}$, $w_a$-unconstrained) are heavily degraded by the geometric degeneracy \cite{Planck:2018vyg}: a broad locus of $(w_0, w_a)$ combinations yields the exact same angular diameter distance to recombination.

Adding \textbf{DESI} BAO measurements (grey contour, $w_0=-0.4\pm0.2$, $w_a=-1.8\pm0.6$) successfully breaks this degeneracy. However, instead of collapsing onto the cosmological constant point, the posterior moves away from $(w_0=-1, w_a=0)$, reflecting the $3.1\sigma$ as reported from \textit{Planck}+\textbf{DESI} \cite{Cortes:2025joz}. Bringing in the full combination with \textbf{PP} supernovae (navy contour, $w_0=-0.85\pm0.06$, $w_a=-0.6\pm0.2$) pulls the posterior back toward $\Lambda\text{CDM}$, though the $w_0 > -1, w_a < 0$ quadrant remains favored at $2.8\sigma$ \cite{DESI:2025zgx}.

Despite this visual displacement of the posterior, the Bayesian evidence tells a different story. Using the \texttt{harmonic} estimator, we find $-\ln\mathcal{Z}_{H,\text{CPL}} = 2165.12 \pm 0.08$ compared to {$-\ln\mathcal{Z}_{H,\Lambda\text{CDM}} = 2162.78 \pm 0.16$}, which yields a Bayes factor of $\ln B_{\Lambda\text{CDM},\text{CPL}} = 2.34 \pm 0.18$.{Accounting for the $0.24$ agreement in $\ln B$ between \texttt{harmonic} and \texttt{UltraNest} established in Sec.~\ref{pure_background}, this reflects \textit{definite} evidence against CPL on the scale of Tab.~\ref{tab:jeffreys_bellido}. We note that for \textit{Planck} alone the Bayes factor is in fact slightly negative, $\ln B_{\Lambda\text{CDM},\text{CPL}} \approx -0.81 \pm 0.16$, i.e.\ CMB data on their own marginally prefer CPL, the preference for $\Lambda$CDM emerges only once the low-redshift geometry is added.} In short, the minor fit improvement provided by CPL is simply not enough to pay for its two additional free parameters.

\begin{figure}[!htbp]
\centering
\includegraphics[width=0.48\textwidth]{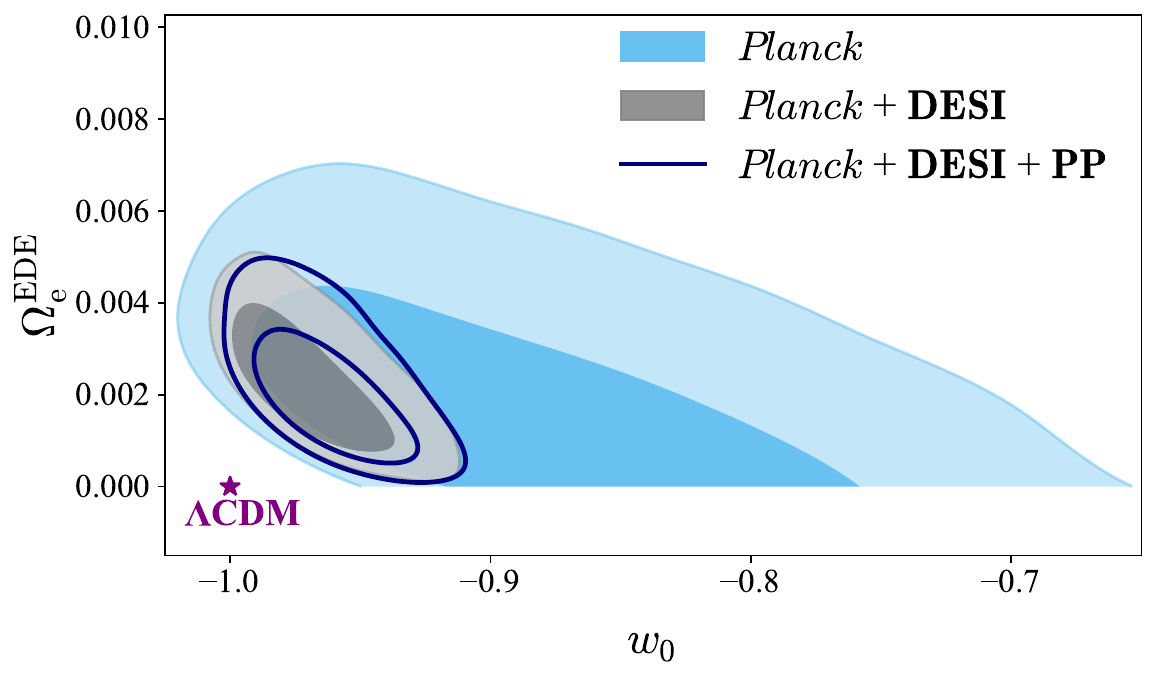}
\caption{Constraints on the EDE parameters $w_0$ and $\Omega^{\rm EDE}_{\rm e}$ at $68\%$ and $95\%$ C.L. The degeneracy present in the \textit{Planck}-only analysis (light blue) is resolved by the inclusion of \textbf{DESI} data (grey). The further addition of \textbf{PP} (navy open contours) yields nested, highly consistent bounds. The purple star marks the exact standard $\Lambda$CDM limit ($w_0 = -1, \Omega^{\rm EDE}_{\rm e} = 0$).}
\label{fig:ede_contours_2D}
\end{figure}

For the EDE model, Fig.~\ref{fig:ede_contours_2D} shows the constraints in the $w_0$--$\Omega^{\rm EDE}_e$ plane. While at 95\% C.L. \textit{Planck} alone permits a long, unconstrained tail toward larger EDE fractions ($10^3\,\Omega_e^{\rm EDE} < 5.3$), combining CMB anisotropies with \textbf{DESI} BAO (grey contour) severely truncates this parameter space. Notably, the inclusion of \textbf{DESI} BAO data plays a decisive role in closing the posterior contours on the EDE fraction for the first time, breaking CMB degeneracies by pinning $w_0$ close to $-1$ ($w_0=-0.97^{+0.01}_{-0.03}$) and tightly capping the EDE fraction at $10^3\,\Omega_e^{\rm EDE} = 2.4^{+0.9}_{-1.0}$. Adding \textbf{PP} supernovae (navy contour) leaves these bounds essentially intact ($10^3\,\Omega_e^{\rm EDE} = 2.0^{+0.6}_{-1.0}$), showing that low-redshift BAO geometry provides the primary anchor. Physically, since $\Omega_{\rm ede}(a) \to \Omega_e^{\rm EDE}$ at early times, this upper bound ($10^3\,\Omega_e^{\rm EDE} = 2.0^{+0.6}_{-1.0}$, or $\Omega_e^{\rm EDE} \lesssim 0.005$ at $95\%$ CL) means that EDE could account for at most $\sim 0.2\%$ of the total energy budget before recombination.

This tight bound leads to a massive Bayesian evidence penalty. The EDE negative log-evidence of $- \ln\mathcal{Z}_{H,\text{EDE}} =  2175.39 \pm 0.04$ translates to a Bayes factor of $\ln B_{\Lambda\text{CDM},\text{EDE}} = 12.61 \pm 0.17$, indicating {\textit{very strong}} evidence in favor of $\Lambda\text{CDM}$. {This heavy penalty is dominated by ``wasted'' prior volume: while the prior allowed $\Omega_e^{\rm EDE} \in [0, 0.5]$, the data restrict it to a tiny slice ($\Omega_e^{\rm EDE} \lesssim 0.005$), penalizing the model for spending prior probability on regions where the likelihood is practically zero. The split can be made explicit through the identity $\ln B_{\Lambda\text{CDM},i} = [\langle\ln\mathcal{L}\rangle_{\Lambda\text{CDM}} - \langle\ln\mathcal{L}\rangle_i] + [D_i - D_{\Lambda\text{CDM}}]$, where $D$ is the Kullback--Leibler divergence of the posterior from the prior: $9.27$ of the total is the volume (Occam) term, while $3.17$ is a genuine degradation of the fit, corresponding to $\Delta\langle\chi^2\rangle=+6.35$ relative to $\Lambda$CDM. The penalty is therefore dominated by wasted prior volume at the $\simeq75\%$ level rather than being entirely due to it. For CPL the same decomposition gives $\ln B = -2.47 + 4.80$: that model \emph{improves} the fit by $\Delta\langle\chi^2\rangle = -4.95$ and is still disfavored, because the volume penalty more than cancels the gain.} Looking beyond dark energy parameters, two notable cosmological implications arise from these full joint fits (Tab.~\ref{tab:results_ede_cpl_comparison}): 

\begin{itemize}

     \item \textbf{Persistence of the Hubble tension:} The inclusion of CMB data, strongly reduces the freedom in the inferred expansion rate. For the full dataset combination, both CPL and EDE converge to nearly identical $h$ values.
     Thus, despite their differences, neither model produces a significant shift in $h$ relative to the $\Lambda$CDM result. \\

    \item \textbf{Bounds on Neutrino mass:} By breaking geometric degeneracies, combining BAO and CMB data significantly tightens the limit on total neutrino mass. For the EDE extension, we find $\sum m_\nu < 0.050 \ \text{eV}$ ($95\%$ CL), whereas CPL allows the considerably weaker bound $\sum m_\nu < 0.13 \ \text{eV}$. The EDE constraint strongly disfavors the inverted mass hierarchy, which requires a minimum sum of $\sum m_\nu \gtrsim 0.10 \ \text{eV}$ from neutrino oscillation experiments \cite{Esteban:2020cvm, Vagnozzi:2017ovm}. {It also lies below the minimum allowed by the normal hierarchy, $\sum m_\nu \gtrsim 0.058 \ \text{eV}$: within this model the posterior places $97.3\%$ of its mass below that threshold. Two caveats temper the statement. First, the $\sum m_\nu$ posterior piles up against the physical boundary at zero, with $41\%$ of its mass at $\sum m_\nu < 0.01 \ \text{eV}$, so the bound is sensitive to the prior placed on a parameter the data do not resolve. Second, our chains model the massive sector as a single massive eigenstate rather than three degenerate ones, a distinction that matters at the percent level precisely near $0.06 \ \text{eV}$. We therefore read this as strongly disfavoring both hierarchies within the EDE model, rather than as an exclusion.} The additional late-time freedom of CPL substantially relaxes the cosmological neutrino mass bound. \\

    \item \textbf{Deviation from $\Lambda$CDM:} The EDE model deviates from the $\Lambda$CDM limit ($w_0=-1$, $\Omega_e^{\rm EDE}=0$) at the $2.0\sigma$ level in both $w_0$ and $\Omega_e^{\rm EDE}$ for the \textit{Planck}+\textbf{DESI}+\textbf{PP} combination, the latter evaluated using the appropriate one-sided uncertainty given the asymmetric $\Omega_e^{\rm EDE}$ posterior. {These two parameters are, however, appreciably anticorrelated, $r(w_0,\Omega_e^{\rm EDE}) = -0.56$, so they cannot be combined as if they were independent \cite{2019PhRvD..99d3506R}. The displacement is therefore a genuine multi-$\sigma$ frequentist effect \cite{2019PhRvD.100d3504H}, which sharpens rather than softens the contrast with the Bayesian evidence.} For the same dataset combination, CPL shows larger deviations from its $\Lambda$CDM limit ($w_0=-1$, $w_a=0$), at approximately $2.5\sigma$ in $w_0$ and $3.0\sigma$ in $w_a$.
\end{itemize}

In summary, the joint CMB analysis reinforces and sharpens our background-only findings: apparent frequentist deviations from $\Lambda\text{CDM}$ dissipate into a Bayesian preference for the minimal model once the extra prior volume is penalized \cite{Ong:2025utx,Ong:2026tta}.

\subsection{Functional reconstruction of the dark energy sector}
\label{Reconstruction}

We reconstruct the functional posteriors of $w_{\rm de}(z)$\footnote{Complementarily, model-agnostic approaches, such as non-parametric reconstructions using Gaussian processes or principal component analysis, aim to directly infer $w(z)$ from data without assuming a specific functional form \cite{Zapata:2025ngr,DESI:2024aqx,Escamilla:2021uoj,Escamilla:2023shf,Zumalacarregui}.} and of the normalized dark energy density $\Omega_{\rm de}(z)\equiv\rho_{\rm fld}(z)/\rho_{\rm crit}(z)$ from the full \textit{Planck}+\textbf{DESI}+\textbf{PP} chains of both models using \texttt{fgivenx}~\citep{fgivenx}.

\begin{figure*}[!ht]
    \centering
    \includegraphics[width=1.0\linewidth]{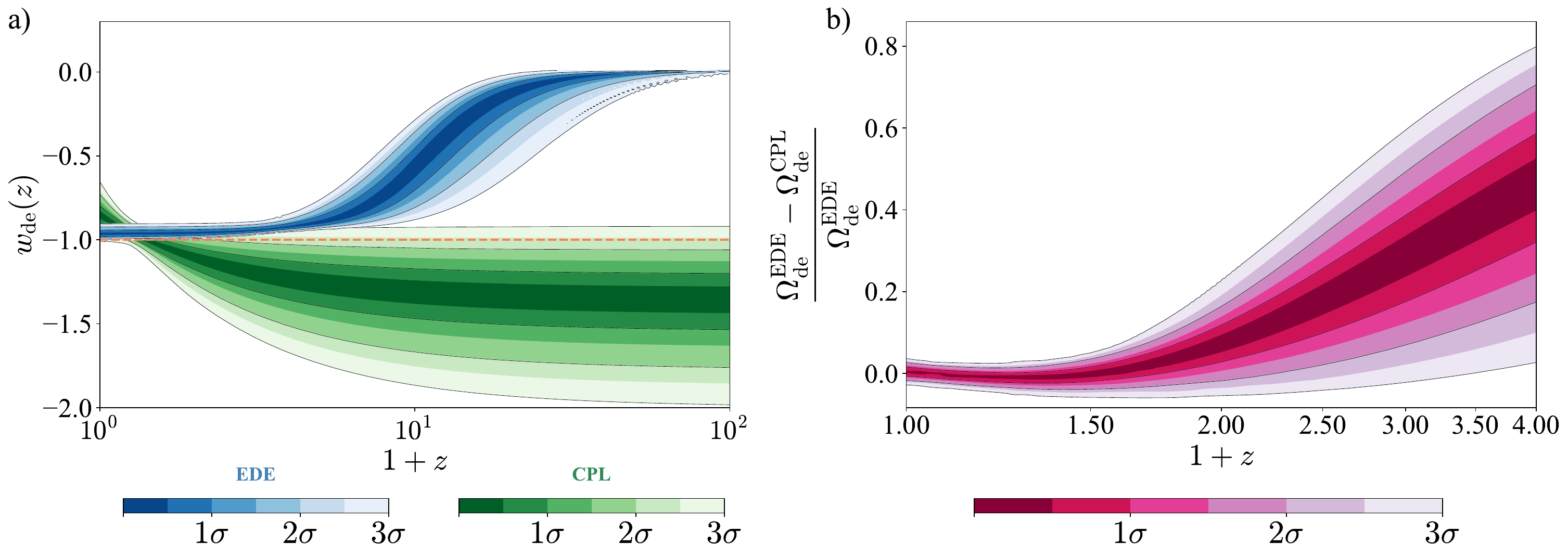}
    \caption{Functional posterior reconstruction of the dark energy sector for
    the EDE and CPL models, from the \textit{Planck}+\textbf{DESI}+\textbf{PP}
    chains via \texttt{fgivenx}, with shaded bands denoting $1\sigma$ -- $3\sigma$
    credible regions. \textbf{a)} Equation of state $w_{\rm de}(z)$ for EDE
    (blue) and CPL (green); both converge to $w_{\rm de} \rightarrow -1$ near $z=0$ but
    depart from one another at higher redshift, EDE toward its matter-tracking
    attractor and CPL toward increasingly phantom values with no physical
    high-redshift limit. \textbf{b)} Fractional residual between the two
    models reconstructed dark energy density,
    $(\Omega_{\rm de}^{\rm EDE}-\Omega_{\rm de}^{\rm CPL})/\Omega_{\rm de}^{\rm EDE}$,
    from paired samples of the two independent chains; the two models agree
    within uncertainties near the present epoch and diverge progressively
    toward higher redshift.}
    \label{fig:functional_reconstruction}
\end{figure*}

In the left panel of Fig.~\ref{fig:functional_reconstruction} we show $w_{\rm de}(z)$ for both models. Near the present epoch both are pinned close to $w_{\rm de}\approx-1$, as expected from the tight low-redshift constraints in Tab.~\ref{tab:results_ede_cpl_comparison}. Their subsequent evolution, however, is qualitatively different: EDE remains in the non-phantom regime, $w_{\rm de}>-1$, whereas CPL crosses the phantom divide, $w_{\rm de}=-1$, and evolves into the phantom region, $w_{\rm de}<-1$. At higher redshift the two diverge for quite different reasons. EDE approaches its early-time attractor and climbs toward $w_{\rm de}\to0$ by $1+z\sim10^2$, tracking the dominant matter component exactly as built into Eq.~\eqref{eq:eos_parameter}. {The faint, thin band that peels off toward larger $1+z$ corresponds to samples with small $\Omega_e^{\rm EDE}$, for which the shape of $w_{\rm de}(z)$ at high redshift is only weakly determined by the data.} CPL has no such attractor: being linear in $(1-a)$, it keeps drifting toward increasingly phantom values as $z$ grows, with the median reaching $w_{\rm de}\sim-1.3$ and the outer band extending past $-2$ by $1+z=10^2$. This is simply what happens when a parametrization built for late-time acceleration is extrapolated well outside the range it was designed for, and it is exactly the kind of behavior a fixed two-parameter fit cannot reveal but a direct functional reconstruction does.

In the right panel of Fig.~\ref{fig:functional_reconstruction}, we illustrate the fractional residual $(\Omega_{\rm de}^{\rm EDE}-\Omega_{\rm de}^{\rm CPL})/\Omega_{\rm de}^{\rm EDE}$, built by pairing samples from the two independent chains. The two models are indistinguishable close to today, with the residual compatible with zero out to $1+z\sim1.5$, which is not surprising since both are fit to reproduce essentially the same low-redshift expansion history. Past that point, the residual grows steadily to a median of $\sim 30$ -- $40\%$ by $1+z=4$, though the $2\sigma$ band remains consistent with zero across most of this range. Specifically, a phantom equation of state dilutes faster into the past than one frozen near $-1$. Consequently, for the same present-day $\Omega_{\rm de,0}$, the CPL reconstructed density falls below the EDE density once the difference in $w_{\rm de}(z)$ shown in Fig.~\ref{fig:functional_reconstruction} spans enough redshift to accumulate.

\section{Conclusions}
\label{Conclusions}

In this work, we have addressed the physics of dark energy under the lens of the latest cosmological dataset combination (\textit{Planck} + \textbf{DESI} + \textbf{PP}), evaluating the tension between early and late modifications to standard $\Lambda\text{CDM}$. By combining robust Bayesian evidence estimation with full likelihood sampling, our analysis provides a clear physical explanation for why early-time solutions face a much steeper observational resistance than late-time dynamical extensions.

Most notably, for the EDE model, \textbf{DESI} BAO data effectively breaks the parameter degeneracies inherent to CMB-only fits. {Whereas \textit{Planck} alone yields only an upper limit} ($10^{3}\,\Omega_{\rm e}^{\rm EDE} < 5.3$), adding \textbf{DESI} closes the likelihood contour and establishes well-defined two-sided bounds ($10^{3}\,\Omega_{\rm e}^{\rm EDE} = 2.4^{+0.9}_{-1.0}$), which tighten further for the full \textit{Planck} + \textbf{DESI} + \textbf{PP} combination ($10^{3}\,\Omega_{\rm e}^{\rm EDE} = 2.0^{+0.6}_{-1.0}$).

To ensure numerical consistency in our evidence calculations, we verified the learned harmonic mean algorithm (\texttt{harmonic}) against direct nested sampling (via \texttt{UltraNest}), {finding that the two estimators agree on $\ln B$ to better than $0.25$ at the background level, in line with established benchmarks in the literature}. By restricting the viable EDE fraction to such a narrow window, less than $1\%$ of its initial prior volume $[0, 0.5]$, the data trigger a severe Occam's razor penalty, resulting in {very strong Bayesian evidence against EDE ($\ln B \approx 12.6$), of which roughly three quarters is wasted prior volume and the remainder a genuine worsening of the fit. In contrast, late-time dynamical dark energy incurs only a modest volume penalty ($\ln B \approx 2.3$), remaining statistically competitive within the standard framework.}

These findings clarify the apparent discrepancy between frequentist metrics and Bayesian model comparison. While frequentist indicators, such as local $\Delta\chi^2$ improvements or two-dimensional contour shifts frequently highlighted in DESI analyses, might suggest a mild preference for dark energy dynamics, these gains do not survive when penalizing for parameter volume. A modest improvement in fit cannot offset the unnatural degree of fine-tuning required by early-time mechanisms once primary CMB data are included. Furthermore, when anchoring EDE models with CMB observations, the capacity to resolve the Hubble tension degrades significantly, yielding $h \approx 0.679$ (\textit{Planck} + \textbf{PP}) and leaving the tension largely intact, while concurrently driving the upper bound on the sum of neutrino masses to an exceptionally strict limit of $\sum m_\nu < 0.05 \ \text{eV}$.

Beyond these headline evidence numbers, the functional reconstruction of Sec.~\ref{Reconstruction} clarifies how each model actually departs from $\Lambda\text{CDM}$ across cosmic history, rather than by how much. Reconstructing $w_{\rm de}(z)$ directly from the chains shows EDE settling onto a bounded, early-time attractor that tracks the dominant matter component, while CPL, once extrapolated beyond the redshift range it was fit to, drifts into an unbounded phantom regime with no physical endpoint. The corresponding reconstruction of $\Omega_{\rm de}(z)$ shows the two models agree within uncertainties near the present epoch but diverge progressively at higher redshift, reaching a median difference of order $30$--$40\%$ by $1+z=4$. This distinction is invisible in the fixed $(w_0,w_a)$ and $(w_0,\Omega_e^{\rm EDE})$ contours of Sec.~\ref{CMB}, yet it is exactly the point this work set out to make: the preference for $\Lambda\text{CDM}$ over both extensions is not simply a matter of degree, but reflects two physically distinct ways a dark energy component can depart from a cosmological constant, one bounded and confined to the early universe, the other open-ended and tied to no particular epoch at all.

To further test these inferences, an immediate priority is the inclusion of small-scale CMB polarization and temperature data from ACT~\cite{AtacamaCosmologyTelescope:2025blo} and SPT~\cite{SPT-3G:2025bzu}. In the coming years, the synergy between next-generation experiments, such as the Simons Observatory~\cite{SimonsObservatory:2018koc} and CMB-S4~\cite{CMB-S4:2016ple} and the final DESI data releases will deliver the sensitivity required to map the dark sector with extreme precision. This upcoming observational baseline will be decisive in determining whether dark energy dynamics represent a true departure from $\Lambda\text{CDM}$ or a spurious statistical fluke, while providing an uncompromising test for any alternative physics in the early universe.

\begin{acknowledgments}
M.A.Z. acknowledges support from the Secretaría de Ciencia, Humanidades, Tecnología e Innovación (SECIHTI) through a studentship. K.C. and G.G.-A. acknowledge the postdoctoral fellowships from SECIHTI.  The authors gratefully acknowledge the computing facilities provided by the Atocatl cluster at LAMOD-UNAM. LAMOD (\url{http://www.lamod.unam.mx/}) is a collaborative effort between the IA, ICN, and IQ institutes at UNAM.
\end{acknowledgments}

\section*{Data Availability}
\phantomsection
\label{data_av}

The data underlying this article will be shared on reasonable request to the corresponding author. 

\appendix
\section{Distances Priors and EDE Impact on the Matter Power Spectrum}
\label{app:A}

In this Appendix, we summarize the priors adopted in our background analysis and illustrate the physical impact of the Early Dark Energy (EDE) fluid on the matter power spectrum $P(k)$. Tab.~\ref{tab:table_priors_bg} specifies the flat top-hat priors assigned to both the baseline $\Lambda\mathrm{CDM}$ cosmological parameters and the extended sector governing EDE and late-time dark energy dynamics ($w_0, w_a, \Omega_e^{\rm EDE}$). These prior ranges are selected to cover all physically viable configurations while preventing boundary effects during nested sampling.
\begin{table}[!htbp]
\caption{Flat top-hat priors imposed on the cosmological parameters for the inference with nested sampling, i.e.\ for the background-only fits of Sec.~\ref{pure_background}. {These ranges are narrower than those of Tab.~\ref{tab:table_priors}, which apply to the MCMC runs including CMB data. The restriction is numerical in origin: the region-construction step of MLFriends becomes unstable when a large fraction of the prior volume returns a vanishing likelihood, as happens when $\Omega_e^{\rm EDE}$ is sampled up to $0.5$ using geometric data alone. Parameters that background probes cannot constrain are therefore confined to physically motivated ranges. Because the posteriors of the CMB-included runs lie entirely within these narrower ranges, the two sets of Bayes factors can be placed on a common footing by rescaling with the ratio of prior volumes of the extra parameters of each extension; this shifts $\ln B_{\Lambda{\rm CDM},{\rm EDE}}$ by $-2.59$ and $\ln B_{\Lambda{\rm CDM},{\rm CPL}}$ by $-1.20$, so that the gap between the two models remains in the range $8.7$--$10.1$ regardless of which prior is adopted.}}
\setlength{\tabcolsep}{10pt}
\renewcommand{\arraystretch}{1.5}
\begin{tabular}{@{}l@{\hspace{3em}}cc@{}}
\toprule
\toprule
Parameter & Prior Range & Distribution \\ 
\midrule
$\omega_{\rm b}$                            & $[0.022, 0.0226]$ & \tophattails  \\
$\omega_{\rm cdm}$                          & $[0.11, 0.13]$    & \tophattails  \\
$h$                                         & $[0.6, 0.8]$      & \tophattails  \\
$\tau_{\rm reio}$                           & {$[0.01, 0.08]$}     & \tophattails  \\
$\ln \left(10^{10} A_{\rm s} \right)$       & $[2.8, 3.2]$      & \tophattails  \\
$\sum m_{\nu}$ [eV]                              & {$[0, 0.1]$}          & \tophattails  \\
$n_s$                                       & $[0.9, 1.0]$      & \tophattails  \\
$w_0$                                       & $[-2.0, -0.5]$    & \tophattails  \\
$\Omega_{\rm e}^{\rm EDE}$                  & $[0.0, 0.1]$      & \tophattails  \\
$w_{\rm a}$                                 & $[-2, 2]$         & \tophattails  \\
\bottomrule
\bottomrule
\end{tabular}
\label{tab:table_priors_bg}
\end{table}

\begin{figure}[!ht]
    \centering
    \includegraphics[width=1.0\linewidth]{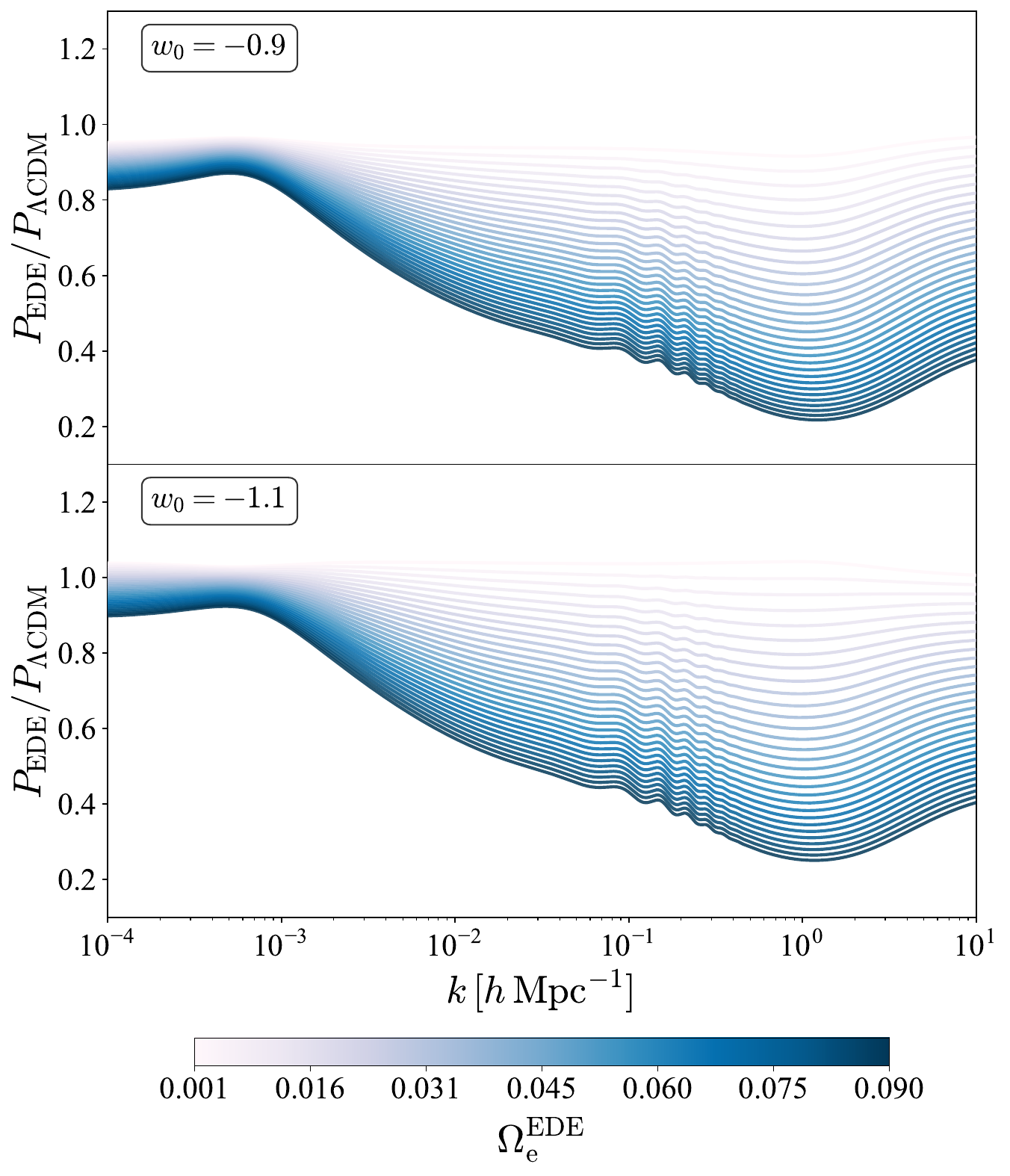}
\caption{Ratio of the matter power spectrum in the EDE model relative to the standard $\Lambda$CDM prediction, defined as $P_{\rm{EDE}}(k)/P_{\Lambda\rm{CDM}}(k)$. The panels display the modifications for two fixed values of the dark energy parameter: $w_0 = -0.9$ (top panel) and $w_0 = -1.1$ (bottom panel). The color gradient represents the fraction of EDE, $\Omega_e^{\text{EDE}}$, showing how increasing the EDE contribution suppresses or enhances structure formation at different scales. The non-linear evolution of the power spectrum at small scales is modeled using the \texttt{Halofit} \cite{2012ApJ...761..152T} approximation.}
\label{fig:matter_pk}
\end{figure}

To isolate the structural modifications introduced by EDE, Fig.~\ref{fig:matter_pk} presents the ratio of the matter power spectrum in the EDE scenario to the standard $\Lambda$CDM baseline, $P_{\rm EDE}(k)/P_{\Lambda\rm CDM}(k)$, as a function of the wavenumber $k$. The presence of an EDE component prior to recombination alters the expansion rate near matter-radiation equality, thereby modifying the sound horizon $r_s$ and changing the growth rate of density perturbations across linear and non-linear scales. The upper panel depicts the response for a quintessence-like late-time equation of state ($w_0 = -0.9$), while the lower panel corresponds to a phantom-like regime ($w_0 = -1.1$). Increasing the fractional EDE density $\Omega_e^{\text{EDE}}$ leads to a progressive suppression of power at intermediate scales, accompanied by characteristic features at small non-linear scales.

\section{Full Parameter Constraints and Posterior Distributions}
\label{app:B}

\begin{figure*}[!ht]
    \centering
    \includegraphics[width=1.0\linewidth]{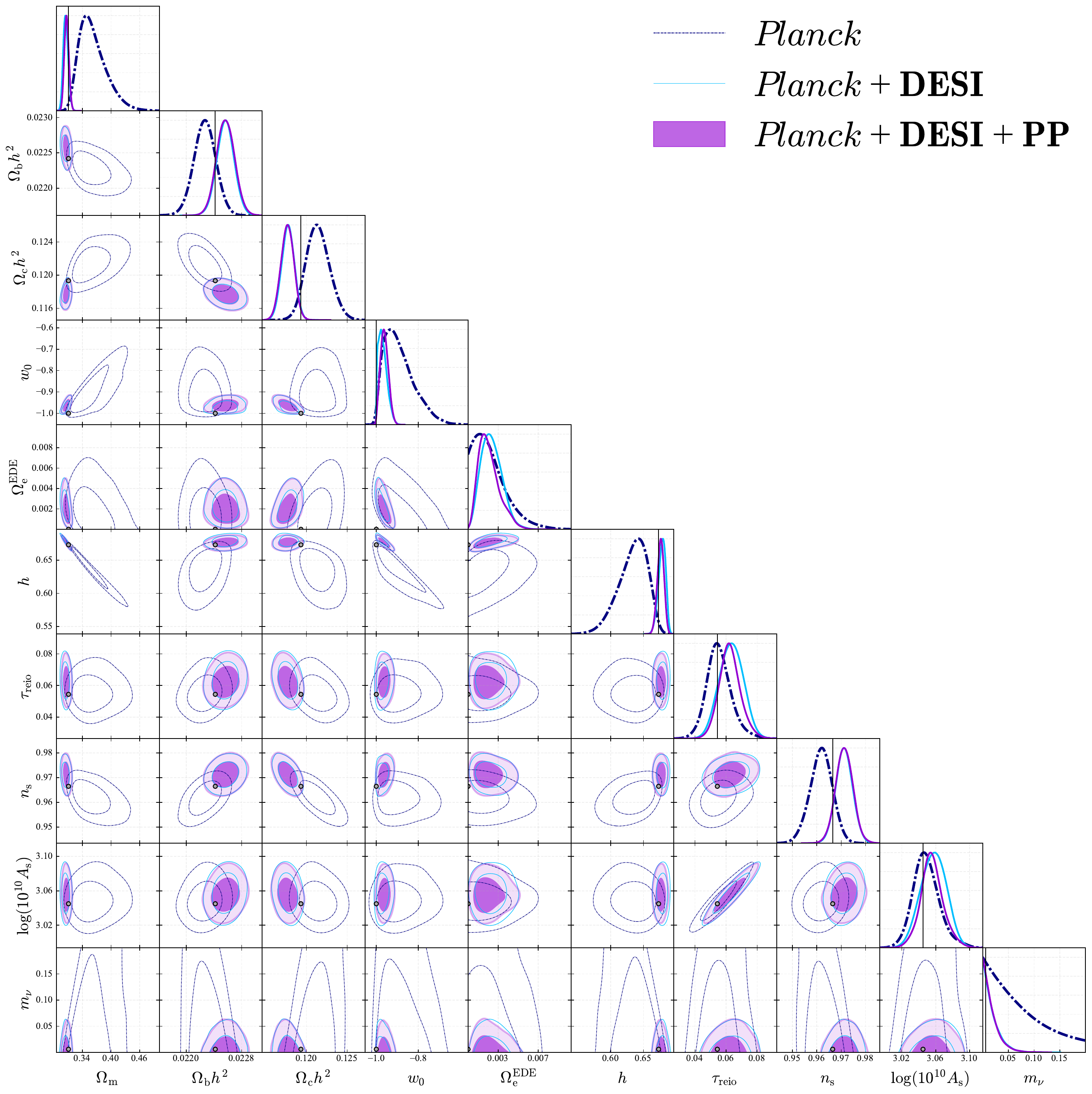}
   \caption{This triangle plot displays the full marginalized posterior distributions and two-dimensional 68\% and 95\% confidence level contours for the key cosmological parameters in the EDE model. We compare the constraints obtained from \textit{Planck} data alone (dark blue), the combination of \textit{Planck} + \textbf{DESI} (light blue), and the full combination of \textit{Planck} + \textbf{DESI} + \textbf{PP} (purple). The black circles represent the best-fit of $\Lambda$CDM.} 
   \label{fig:triangle_all}
\end{figure*}

In Fig.~\ref{fig:triangle_all}, we present the full multi-dimensional posterior distributions and two-dimensional joint confidence regions (68\% and 95\% C.L.) for all primary and derived parameters of the EDE model. Comparing the constraints derived from \textit{Planck} CMB data alone (dark blue) to those including \textbf{DESI} BAO measurements (light blue) and the combined \textit{Planck} + \textbf{DESI} + \textbf{PP} dataset (purple) illustrates the progressive breaking of geometric degeneracies, demonstrating how late-time probes pull the parameter posteriors relative to the standard $\Lambda$CDM best-fit values.

\nocite{*}

\clearpage

\bibliographystyle{utphys}
\bibliography{my_bib}

@article{Escamilla:2023shf,
    author = "Escamilla, Luis A. and Akarsu, Ozgur and Di Valentino, Eleonora and Vazquez, J. Alberto",
    title = "{Model-independent reconstruction of the interacting dark energy kernel: Binned and Gaussian process}",
    eprint = "2305.16290",
    archivePrefix = "arXiv",
    primaryClass = "astro-ph.CO",
    doi = "10.1088/1475-7516/2023/11/051",
    journal = "JCAP",
    volume = "11",
    pages = "051",
    year = "2023"
}

@article{Liddle:2007fy,
    author = "Liddle, Andrew R",
    title = "{Information criteria for astrophysical model selection}",
    eprint = "astro-ph/0701113",
    archivePrefix = "arXiv",
    doi = "10.1111/j.1745-3933.2007.00306.x",
    journal = "Mon. Not. Roy. Astron. Soc.",
    volume = "377",
    pages = "L74--L78",
    year = "2007"
}

@article{Hu:2008zd,
    author = "Hu, Wayne",
    title = "{Parametrized Post-Friedmann Signatures of Acceleration in the CMB}",
    eprint = "0801.2433",
    archivePrefix = "arXiv",
    primaryClass = "astro-ph",
    doi = "10.1103/PhysRevD.77.103524",
    journal = "Phys. Rev. D",
    volume = "77",
    pages = "103524",
    year = "2008"
}

@article{Hu:2007pj,
    author = "Hu, Wayne and Sawicki, Ignacy",
    title = "{A Parameterized Post-Friedmann Framework for Modified Gravity}",
    eprint = "0708.1190",
    archivePrefix = "arXiv",
    primaryClass = "astro-ph",
    doi = "10.1103/PhysRevD.76.104043",
    journal = "Phys. Rev. D",
    volume = "76",
    pages = "104043",
    year = "2007"
}

@article{Fang:2008sn,
    author = "Fang, Wenjuan and Hu, Wayne and Lewis, Antony",
    title = "{Crossing the Phantom Divide with Parameterized Post-Friedmann Dark Energy}",
    eprint = "0808.3125",
    archivePrefix = "arXiv",
    primaryClass = "astro-ph",
    doi = "10.1103/PhysRevD.78.087303",
    journal = "Phys. Rev. D",
    volume = "78",
    pages = "087303",
    year = "2008"
}

@article{Lewis:2002ah,
    author = "Lewis, Antony and Bridle, Sarah",
    title = "{Cosmological parameters from CMB and other data: A Monte Carlo approach}",
    eprint = "astro-ph/0205436",
    archivePrefix = "arXiv",
    doi = "10.1103/PhysRevD.66.103511",
    journal = "Phys. Rev. D",
    volume = "66",
    pages = "103511",
    year = "2002"
}

@article{DESI:2025fii,
    author = "Lodha, K. and others",
    collaboration = "DESI",
    title = "{Extended dark energy analysis using DESI DR2 BAO measurements}",
    eprint = "2503.14743",
    archivePrefix = "arXiv",
    primaryClass = "astro-ph.CO",
    reportNumber = "FERMILAB-PUB-25-0164-PPD",
    doi = "10.1103/w4c6-1r5j",
    journal = "Phys. Rev. D",
    volume = "112",
    number = "8",
    pages = "083511",
    year = "2025"
}

@ARTICLE{2020PhRvD.102d3507H,
       author = {{Hill}, J. Colin and {McDonough}, Evan and {Toomey}, Michael W. and {Alexander}, Stephon},
        title = "{Early dark energy does not restore cosmological concordance}",
      journal = {\prd},
         year = 2020,
        month = aug,
       volume = {102},
       number = {4},
          eid = {043507},
        pages = {043507},
          doi = {10.1103/PhysRevD.102.043507},
archivePrefix = {arXiv},
       eprint = {2003.07355},
 primaryClass = {astro-ph.CO},
       adsurl = {https://ui.adsabs.harvard.edu/abs/2020PhRvD.102d3507H}
}

@ARTICLE{2019PhRvD..99d3506R,
       author = {{Raveri}, Marco and {Hu}, Wayne},
        title = "{Concordance and discordance in cosmology}",
      journal = {\prd},
         year = 2019,
        month = 02,
       volume = {99},
       number = {4},
          eid = {043506},
        pages = {043506},
          doi = {10.1103/PhysRevD.99.043506},
archivePrefix = {arXiv},
       eprint = {1806.04649},
 primaryClass = {astro-ph.CO},
       adsurl = {https://ui.adsabs.harvard.edu/abs/2019PhRvD..99d3506R}
}

@ARTICLE{2020A&A...641A...8P,
    author = "Aghanim, N. and others",
    collaboration = "Planck",
    title = "{Planck 2018 results. VIII. Gravitational lensing}",
    eprint = "1807.06210",
    archivePrefix = "arXiv",
    primaryClass = "astro-ph.CO",
    doi = "10.1051/0004-6361/201833886",
    journal = "Astron. Astrophys.",
    volume = "641",
    pages = "A8",
    year = "2020"
}

@ARTICLE{2026NIMPA108270857B,
       author = {{Barlow}, Roger and {Brazzale}, Alessandra Rosalba and {Volobouev}, Igor},
        title = "{Asymmetric errors}",
      journal = {Nuclear Instruments and Methods in Physics Research A},
         year = 2026,
        month = feb,
       volume = {1082},
          eid = {170857},
        pages = {170857},
          doi = {10.1016/j.nima.2025.170857},
archivePrefix = {arXiv},
       eprint = {2411.15499},
 primaryClass = {stat.ME},
       adsurl = {https://ui.adsabs.harvard.edu/abs/2026NIMPA108270857B}
}

@article{fgivenx,
    doi = {10.21105/joss.00849},
    url = {http://dx.doi.org/10.21105/joss.00849},
    year  = {2018},
    month = {Aug},
    publisher = {The Open Journal},
    volume = {3},
    number = {28},
    author = {Will Handley},
    title = {fgivenx: Functional Posterior Plotter},
    journal = {The Journal of Open Source Software}
}

@article{Newton1994,
 ISSN = {00359246},
 URL = {http://www.jstor.org/stable/2346025},
 author = {Michael A. Newton and Adrian E. Raftery},
 journal = {Journal of the Royal Statistical Society. Series B (Methodological)},
 number = {1},
 pages = {3--48},
 publisher = {[Royal Statistical Society, Oxford University Press]},
 title = {Approximate Bayesian Inference with the Weighted Likelihood Bootstrap},
 urldate = {2026-09-03},
 volume = {56},
 year = {1994}
}

@article{Pantazis:2016nky,
    author = "Pantazis, G. and Nesseris, S. and Perivolaropoulos, L.",
    title = "{Comparison of thawing and freezing dark energy parametrizations}",
    eprint = "1603.02164",
    archivePrefix = "arXiv",
    primaryClass = "astro-ph.CO",
    reportNumber = "IFT-UAM-CSIC-16-023",
    doi = "10.1103/PhysRevD.93.103503",
    journal = "Phys. Rev. D",
    volume = "93",
    number = "10",
    pages = "103503",
    year = "2016"
}

@article{CosmoVerseNetwork:2025alb,
    author = "Di Valentino, Eleonora and others",
    collaboration = "CosmoVerse Network",
    title = "{The CosmoVerse White Paper: Addressing observational tensions in cosmology with systematics and fundamental physics}",
    eprint = "2504.01669",
    archivePrefix = "arXiv",
    primaryClass = "astro-ph.CO",
    doi = "10.1016/j.dark.2025.101965",
    journal = "Phys. Dark Univ.",
    volume = "49",
    pages = "101965",
    year = "2025"
}

@article{Albrecht:1999rm,
    author = "Albrecht, Andreas and Skordis, Constantinos",
    title = "{Phenomenology of a realistic accelerating universe using only Planck scale physics}",
    eprint = "astro-ph/9908085",
    archivePrefix = "arXiv",
    doi = "10.1103/PhysRevLett.84.2076",
    journal = "Phys. Rev. Lett.",
    volume = "84",
    pages = "2076--2079",
    year = "2000"
}

@article{Ong:2025utx,
    author = "Ong, Dily Duan Yi and Yallup, David and Handley, Will",
    title = "{A Bayesian Perspective on Evidence for Evolving Dark Energy}",
    eprint = "2511.10631",
    archivePrefix = "arXiv",
    primaryClass = "astro-ph.CO",
    month = "11",
    year = "2025"
}

@article{Adil:2026kfn,
    author = {Adil, Shahnawaz A. and Zapata, Miguel A. and Akarsu, {\"O}zg{\"u}r and Vazquez, J. Alberto},
    title = "{Background-level reconstruction of scalar-field potentials from dark-energy histories and comparison with analytic potential families}",
    journal = "arXiv e-prints",
    eprint = "2603.14693",
    archivePrefix = "arXiv",
    primaryClass = "astro-ph.CO",
    month = "3",
    year = "2026"
}

@article{Trotta:2008qt,
    author = "Trotta, Roberto",
    title = "{Bayes in the sky: Bayesian inference and model selection in cosmology}",
    eprint = "0803.4089",
    archivePrefix = "arXiv",
    primaryClass = "astro-ph",
    doi = "10.1080/00107510802066753",
    journal = "Contemp. Phys.",
    volume = "49",
    pages = "71--104",
    year = "2008"
}

@ARTICLE{2023ARNPS..73..153K,
       author = {{Kamionkowski}, Marc and {Riess}, Adam G.},
        title = "{The Hubble Tension and Early Dark Energy}",
      journal = {Annual Review of Nuclear and Particle Science},
         year = 2023,
        month = sep,
       volume = {73},
        pages = {153-180},
          doi = {10.1146/annurev-nucl-111422-024107},
archivePrefix = {arXiv},
       eprint = {2211.04492},
 primaryClass = {astro-ph.CO},
       adsurl = {https://ui.adsabs.harvard.edu/abs/2023ARNPS..73..153K}
}

@ARTICLE{2019PhRvD.100d3504H,
       author = {{Handley}, Will and {Lemos}, Pablo},
        title = "{Quantifying tensions in cosmological parameters: Interpreting the DES evidence ratio}",
      journal = {\prd},
         year = 2019,
        month = aug,
       volume = {100},
       number = {4},
          eid = {043504},
        pages = {043504},
          doi = {10.1103/PhysRevD.100.043504},
archivePrefix = {arXiv},
       eprint = {1902.04029},
 primaryClass = {astro-ph.CO},
       adsurl = {https://ui.adsabs.harvard.edu/abs/2019PhRvD.100d3504H}
}

@ARTICLE{2023PDU....4201348P,
       author = {{Poulin}, Vivian and {Smith}, Tristan L. and {Karwal}, Tanvi},
        title = "{The Ups and Downs of Early Dark Energy solutions to the Hubble tension: A review of models, hints and constraints circa 2023}",
      journal = {Physics of the Dark Universe},
         year = 2023,
        month = dec,
       volume = {42},
          eid = {101348},
        pages = {101348},
          doi = {10.1016/j.dark.2023.101348},
archivePrefix = {arXiv},
       eprint = {2302.09032},
 primaryClass = {astro-ph.CO},
       adsurl = {https://ui.adsabs.harvard.edu/abs/2023PDU....4201348P}
}

@ARTICLE{2025PDU....4801902J,
     author = {{Jiang}, Jun-Qian},
     title={Status of early dark energy after DESI: the role of {$\Omega_m$} and {$r_s H_0$}},
     journal = {Physics of the Dark Universe},
         year = 2025,
        month = may,
       volume = {48},
          eid = {101902},
        pages = {101902},
          doi ={10.1016/j.dark.2025.101902},
archivePrefix = {arXiv},
       eprint = {2502.15541},
 primaryClass = {astro-ph.CO},
       adsurl = {https://ui.adsabs.harvard.edu/abs/2025PDU....4801902J}
}

@ARTICLE{2023PhRvD.108d3513H,
       author = {{Herold}, Laura and {Ferreira}, Elisa G.~M.},
        title = "{Resolving the Hubble tension with early dark energy}",
      journal = {\prd},
         year = 2023,
        month = aug,
       volume = {108},
       number = {4},
          eid = {043513},
        pages = {043513},
          doi = {10.1103/PhysRevD.108.043513},
archivePrefix = {arXiv},
       eprint = {2210.16296},
 primaryClass = {astro-ph.CO},
       adsurl = {https://ui.adsabs.harvard.edu/abs/2023PhRvD.108d3513H}
}

@ARTICLE{2024IJMPD..3330003M,
       author = {{McDonough}, Evan and {Colin Hill}, J. and {Ivanov}, Mikhail M. and {La Posta}, Adrien and {Toomey}, Michael W.},
        title = "{Observational constraints on early dark energy}",
      journal = {International Journal of Modern Physics D},
         year = 2024,
        month = jan,
       volume = {33},
       number = {11},
          eid = {2430003},
        pages = {2430003},
          doi = {10.1142/S0218271824300039},
archivePrefix = {arXiv},
       eprint = {2310.19899},
 primaryClass = {astro-ph.CO},
       adsurl = {https://ui.adsabs.harvard.edu/abs/2024IJMPD..3330003M}
}

@article{DESI:2025zpo,
    author = "Abdul Karim, M. and others",
    collaboration = "DESI",
    title = "{DESI DR2 results. I. Baryon acoustic oscillations from the Lyman alpha forest}",
    eprint = "2503.14739",
    archivePrefix = "arXiv",
    primaryClass = "astro-ph.CO",
    reportNumber = "FERMILAB-PUB-25-0167-PPD",
    doi = "10.1103/2wwn-xjm5",
    journal = "Phys. Rev. D",
    volume = "112",
    number = "8",
    pages = "083514",
    year = "2025"
}

@inproceedings{Peebles:2024txt,
    author = "Peebles, P. J. E.",
    title = "{Status of the $\Lambda$CDM theory: supporting evidence and anomalies}",
    eprint = "2405.18307",
    archivePrefix = "arXiv",
    primaryClass = "astro-ph.CO",
    month = "5",
    year = "2024"
}

@article{DESI:2025gwf,
    author = "Garcia-Quintero, C. and others",
    collaboration = "DESI",
    title = "{Cosmological implications of DESI DR2 BAO measurements in light of the latest ACT DR6 CMB data}",
    eprint = "2504.18464",
    archivePrefix = "arXiv",
    primaryClass = "astro-ph.CO",
    reportNumber = "FERMILAB-PUB-25-0281-PPD",
    doi = "10.1103/d6yc-xpqb",
    journal = "Phys. Rev. D",
    volume = "112",
    number = "8",
    pages = "083529",
    year = "2025"
}

@article{Zapata:2025ngr,
    author = "Zapata, Miguel A. and Garcia-Arroyo, Gabriela and Adil, Shahnawaz A. and Vazquez, J. Alberto",
    title = "{How holographic is the dark energy? A spline nodal reconstruction approach}",
    eprint = "2507.22292",
    archivePrefix = "arXiv",
    primaryClass = "astro-ph.CO",
    doi = "10.1088/1475-7516/2026/05/058",
    journal = "JCAP",
    volume = "05",
    pages = "058",
    year = "2026"
}

@article{Zumalacarregui,
  title = {Reconstructing the dark energy density in light of DESI BAO observations},
  author = {Berti, Maria and Bellini, Emilio and Bonvin, Camille and Kunz, Martin and Viel, Matteo and Zumalacarregui, Miguel},
  journal = {Phys. Rev. D},
  volume = {112},
  issue = {2},
  pages = {023518},
  numpages = {15},
  year = {2025},
  month = {Jul},
  publisher = {American Physical Society},
  doi = {10.1103/dj3k-84v4},
  url = {https://link.aps.org/doi/10.1103/dj3k-84v4}
}

@article{Escamilla:2021uoj,
    author = "Escamilla, Luis A. and Vazquez, J. Alberto",
    title = "{Model selection applied to reconstructions of the Dark Energy}",
    eprint = "2111.10457",
    archivePrefix = "arXiv",
    primaryClass = "astro-ph.CO",
    doi = "10.1140/epjc/s10052-023-11404-2",
    journal = "Eur. Phys. J. C",
    volume = "83",
    number = "3",
    pages = "251",
    year = "2023"
}

@article{DESI:2024aqx,
    author = "Calderon, R. and others",
    collaboration = "DESI",
    title = "{DESI 2024: reconstructing dark energy using crossing statistics with DESI DR1 BAO data}",
    eprint = "2405.04216",
    archivePrefix = "arXiv",
    primaryClass = "astro-ph.CO",
    doi = "10.1088/1475-7516/2024/10/048",
    journal = "JCAP",
    volume = "10",
    pages = "048",
    year = "2024"
}

@article{Linder,
  title = {Exploring the Expansion History of the Universe},
  author = {Linder, Eric V.},
  journal = {Phys. Rev. Lett.},
  volume = {90},
  issue = {9},
  pages = {091301},
  numpages = {4},
  year = {2003},
  month = {Mar},
  publisher = {American Physical Society},
  doi = {10.1103/PhysRevLett.90.091301},
  url = {https://link.aps.org/doi/10.1103/PhysRevLett.90.091301}
}

@article{CHEVALLIER_2001,
   title={ACCELERATING UNIVERSES WITH SCALING DARK MATTER},
   volume={10},
   ISSN={1793-6594},
   url={http://dx.doi.org/10.1142/S0218271801000822},
   DOI={10.1142/s0218271801000822},
   number={02},
   journal={International Journal of Modern Physics D},
   publisher={World Scientific Pub Co Pte Lt},
   author={CHEVALLIER, MICHEL and POLARSKI, DAVID},
   year={2001},
   month=apr, pages={213–223} }

@ARTICLE{2022PDU....3601037R,
       author = {{Roy}, Nandan and {Goswami}, Sangita and {Das}, Sudipta},
        title = "{Quintessence or phantom: Study of scalar field dark energy models through a general parametrization of the Hubble parameter}",
      journal = {Physics of the Dark Universe},
         year = 2022,
        month = jun,
       volume = {36},
          eid = {101037},
        pages = {101037},
          doi = {10.1016/j.dark.2022.101037},
archivePrefix = {arXiv},
       eprint = {2201.09306},
 primaryClass = {astro-ph.CO},
       adsurl = {https://ui.adsabs.harvard.edu/abs/2022PDU....3601037R}
}

@article{Garcia-Arroyo:2024tqq,
    author = "Garcia-Arroyo, Gabriela and Ure{\~n}a-L{\'o}pez, L. Arturo and V{\'a}zquez, J. Alberto",
    title = "{Interacting scalar fields: Dark matter and early dark energy}",
    eprint = "2402.08815",
    archivePrefix = "arXiv",
    primaryClass = "astro-ph.CO",
    doi = "10.1103/PhysRevD.110.023529",
    journal = "Phys. Rev. D",
    volume = "110",
    number = "2",
    pages = "023529",
    year = "2024"
}

@ARTICLE{2023Univ....9...94H,
       author = {{Hu}, Jian-Ping and {Wang}, Fa-Yin},
        title = "{Hubble Tension: The Evidence of New Physics}",
      journal = {Universe},
         year = 2023,
        month = feb,
       volume = {9},
       number = {2},
          eid = {94},
        pages = {94},
          doi = {10.3390/universe9020094},
archivePrefix = {arXiv},
       eprint = {2302.05709},
 primaryClass = {astro-ph.CO},
       adsurl = {https://ui.adsabs.harvard.edu/abs/2023Univ....9...94H}
}

@ARTICLE{2013PhRvD..87h3009P,
       author = {{Pettorino}, Valeria and {Amendola}, Luca and {Wetterich}, Christof},
        title = "{How early is early dark energy?}",
      journal = {\prd},
         year = 2013,
        month = apr,
       volume = {87},
       number = {8},
          eid = {083009},
        pages = {083009},
          doi = {10.1103/PhysRevD.87.083009},
archivePrefix = {arXiv},
       eprint = {1301.5279},
 primaryClass = {astro-ph.CO},
       adsurl = {https://ui.adsabs.harvard.edu/abs/2013PhRvD..87h3009P}
}

@article{Niedermann:2019olb,
    author = "Niedermann, Florian and Sloth, Martin S.",
    title = "{New early dark energy}",
    eprint = "1910.10739",
    archivePrefix = "arXiv",
    primaryClass = "astro-ph.CO",
    doi = "10.1103/PhysRevD.103.L041303",
    journal = "Phys. Rev. D",
    volume = "103",
    number = "4",
    pages = "L041303",
    year = "2021"
}

@article{Sabla:2022xzj,
    author = "Sabla, Vivian I. and Caldwell, Robert R.",
    title = "{Microphysics of early dark energy}",
    eprint = "2202.08291",
    archivePrefix = "arXiv",
    primaryClass = "astro-ph.CO",
    doi = "10.1103/PhysRevD.106.063526",
    journal = "Phys. Rev. D",
    volume = "106",
    number = "6",
    pages = "063526",
    year = "2022"
}

@article{Poulin:2018cxd,
    author = "Poulin, Vivian and Smith, Tristan L. and Karwal, Tanvi and Kamionkowski, Marc",
    title = "{Early Dark Energy Can Resolve The Hubble Tension}",
    eprint = "1811.04083",
    archivePrefix = "arXiv",
    primaryClass = "astro-ph.CO",
    doi = "10.1103/PhysRevLett.122.221301",
    journal = "Phys. Rev. Lett.",
    volume = "122",
    number = "22",
    pages = "221301",
    year = "2019"
}

@article{Lin:2025xbw,
    author = "Lin, Kiyam and Polanska, Alicja and Piras, Davide and Spurio Mancini, Alessio and McEwen, Jason D.",
    title = "{Savage-Dickey density ratio estimation with normalizing flows for Bayesian model comparison}",
    eprint = "2506.04339",
    archivePrefix = "arXiv",
    primaryClass = "astro-ph.CO",
    month = "6",
    year = "2025"
}

@article{Esteban:2020cvm,
    author = "Esteban, Ivan and Gonzalez-Garcia, M. C. and Maltoni, Michele and Schwetz, Thomas and Zhou, Albert",
    title = "{The fate of hints: updated global analysis of three-flavor neutrino oscillations}",
    eprint = "2007.14792",
    archivePrefix = "arXiv",
    primaryClass = "hep-ph",
    reportNumber = "IFT-UAM/CSIC-112, YITP-SB-2020-21",
    doi = "10.1007/JHEP09(2020)178",
    journal = "JHEP",
    volume = "09",
    pages = "178",
    year = "2020"
}

@article{Vagnozzi:2017ovm,
    author = "Vagnozzi, Sunny and Giusarma, Elena and Mena, Olga and Freese, Katherine and Gerbino, Martina and Ho, Shirley and Lattanzi, Massimiliano",
    title = "{Unveiling $\nu$ secrets with cosmological data: neutrino masses and mass hierarchy}",
    eprint = "1701.08172",
    archivePrefix = "arXiv",
    primaryClass = "astro-ph.CO",
    doi = "10.1103/PhysRevD.96.123503",
    journal = "Phys. Rev. D",
    volume = "96",
    number = "12",
    pages = "123503",
    year = "2017"
}

@article{Ong:2026tta,
    author = "Ong, Dily Duan Yi and Yallup, David and Handley, Will",
    title = "{The Bayesian view of DESI DR2 with unimpeded: Evidence and tension in a combined analysis with CMB and supernovae across cosmological models}",
    eprint = "2603.05472",
    archivePrefix = "arXiv",
    primaryClass = "astro-ph.CO",
    month = "3",
    year = "2026"
}

@article{Piras:2024dml,
    author = "Piras, Davide and Polanska, Alicja and Spurio Mancini, Alessio and Price, Matthew A. and McEwen, Jason D.",
    title = "{The future of cosmological likelihood-based inference: accelerated high-dimensional parameter estimation and model comparison}",
    eprint = "2405.12965",
    archivePrefix = "arXiv",
    primaryClass = "astro-ph.CO",
    doi = "10.33232/001c.123368",
    month = "5",
    year = "2024"
}

@article{Carrion:2024jur,
    author = "Carrion, Karim and Spurio Mancini, Alessio and Piras, Davide and Hidalgo, Juan Carlos",
    title = "{Testing interacting dark energy with Stage IV cosmic shear surveys through differentiable neural emulators}",
    eprint = "2410.10603",
    archivePrefix = "arXiv",
    primaryClass = "astro-ph.CO",
    doi = "10.1093/mnras/staf663",
    journal = "Mon. Not. Roy. Astron. Soc.",
    volume = "539",
    number = "4",
    pages = "3220--3228",
    year = "2025"
}

@article{Abdalla:2022yfr,
    author = "Abdalla, Elcio and others",
    title = "{Cosmology intertwined: A review of the particle physics, astrophysics, and cosmology associated with the cosmological tensions and anomalies}",
    eprint = "2203.06142",
    archivePrefix = "arXiv",
    primaryClass = "astro-ph.CO",
    reportNumber = "FERMILAB-CONF-22-192-SCD",
    doi = "10.1016/j.jheap.2022.04.002",
    journal = "JHEAp",
    volume = "34",
    pages = "49--211",
    year = "2022"
}

@article{LINDER200616,
    author = "Linder, Eric V.",
    title = "{Dark Energy in the Dark Ages}",
    eprint = "astro-ph/0603584",
    archivePrefix = "arXiv",
    doi = "10.1016/j.astropartphys.2006.04.004",
    journal = "Astropart. Phys.",
    volume = "26",
    pages = "16--21",
    year = "2006"
}

@article{Torrado:2020dgo,
    author = "Torrado, Jesus and Lewis, Antony",
    title = "{Cobaya: Code for Bayesian Analysis of hierarchical physical models}",
    eprint = "2005.05290",
    archivePrefix = "arXiv",
    primaryClass = "astro-ph.IM",
    reportNumber = "TTK-20-15",
    doi = "10.1088/1475-7516/2021/05/057",
    journal = "JCAP",
    volume = "05",
    pages = "057",
    year = "2021"
}

@ARTICLE{2011arXiv1104.2932L,
       author = {{Lesgourgues}, Julien},
        title = "{The Cosmic Linear Anisotropy Solving System (CLASS) I: Overview}",
      journal = {arXiv e-prints},
         year = 2011,
        month = apr,
          eid = {arXiv:1104.2932},
        pages = {arXiv:1104.2932},
          doi = {10.48550/arXiv.1104.2932},
archivePrefix = {arXiv},
       eprint = {1104.2932},
 primaryClass = {astro-ph.IM},
       adsurl = {https://ui.adsabs.harvard.edu/abs/2011arXiv1104.2932L}
}

@ARTICLE{2011JCAP...07..034B,
       author = {{Blas}, Diego and {Lesgourgues}, Julien and {Tram}, Thomas},
        title = "{The Cosmic Linear Anisotropy Solving System (CLASS). Part II: Approximation schemes}",
      journal = {\jcap},
         year = 2011,
        month = jul,
       volume = {2011},
       number = {7},
          eid = {034},
        pages = {034},
          doi = {10.1088/1475-7516/2011/07/034},
archivePrefix = {arXiv},
       eprint = {1104.2933},
 primaryClass = {astro-ph.CO},
       adsurl = {https://ui.adsabs.harvard.edu/abs/2011JCAP...07..034B}
}

@ARTICLE{2016S&C....26..383B,
       author = {{Buchner}, Johannes},
        title = "{A statistical test for Nested Sampling algorithms}",
      journal = {Statistics and Computing},
         year = 2016,
        month = jan,
       volume = {26},
       number = {1-2},
        pages = {383-392},
          doi = {10.1007/s11222-014-9512-y},
archivePrefix = {arXiv},
       eprint = {1407.5459},
 primaryClass = {stat.CO},
       adsurl = {https://ui.adsabs.harvard.edu/abs/2016S&C....26..383B}
}

@ARTICLE{2019PASP..131j8005B,
       author = {{Buchner}, Johannes},
        title = "{Collaborative Nested Sampling: Big Data versus Complex Physical Models}",
      journal = {\pasp},
         year = 2019,
        month = oct,
       volume = {131},
       number = {1004},
        pages = {108005},
          doi = {10.1088/1538-3873/aae7fc},
archivePrefix = {arXiv},
       eprint = {1707.04476},
 primaryClass = {stat.CO},
       adsurl = {https://ui.adsabs.harvard.edu/abs/2019PASP..131j8005B}
}

@ARTICLE{2021JOSS....6.3001B,
       author = {{Buchner}, Johannes},
        title = "{UltraNest - a robust, general purpose Bayesian inference engine}",
      journal = {The Journal of Open Source Software},
         year = 2021,
        month = apr,
       volume = {6},
       number = {60},
          eid = {3001},
        pages = {3001},
          doi = {10.21105/joss.03001},
archivePrefix = {arXiv},
       eprint = {2101.09604},
 primaryClass = {stat.CO},
       adsurl = {https://ui.adsabs.harvard.edu/abs/2021JOSS....6.3001B}
}

@book{jeffreys1998theory,
  title={The theory of probability},
  author={Jeffreys, Harold},
  year={1998},
  publisher={OuP Oxford}
}

@article{Nesseris:2012cq,
    author = "Nesseris, Savvas and Garcia-Bellido, Juan",
    title = "{Is the Jeffreys' scale a reliable tool for Bayesian model comparison in cosmology?}",
    eprint = "1210.7652",
    archivePrefix = "arXiv",
    primaryClass = "astro-ph.CO",
    reportNumber = "IFT-UAM-CSIC-12-95",
    doi = "10.1088/1475-7516/2013/08/036",
    journal = "JCAP",
    volume = "08",
    pages = "036",
    year = "2013"
}

@article{harmonic,
   author = {Jason~D.~McEwen and Christopher~G.~R.~Wallis and Matthew~A.~Price and Matthew~M.~Docherty},
    title = {Machine learning assisted {B}ayesian model comparison: learnt harmonic mean estimator},
  journal = {ArXiv},
   eprint = {arXiv:2111.12720},
     year = 2021
}

@misc{polanska2024learned,
    title={Learned harmonic mean estimation of the Bayesian evidence with normalizing flows},
    author={Alicja Polanska and Matthew A. Price and Davide Piras and Alessio Spurio Mancini and Jason D. McEwen},
    year={2024},
    eprint={2405.05969},
    archivePrefix={arXiv},
    primaryClass={astro-ph.IM}
}

@ARTICLE{1995ApJ...455....7M,
       author = {{Ma}, Chung-Pei and {Bertschinger}, Edmund},
        title = "{Cosmological Perturbation Theory in the Synchronous and Conformal Newtonian Gauges}",
      journal = {\apj},
         year = 1995,
        month = dec,
       volume = {455},
        pages = {7},
          doi = {10.1086/176550},
archivePrefix = {arXiv},
       eprint = {astro-ph/9506072},
 primaryClass = {astro-ph},
       adsurl = {https://ui.adsabs.harvard.edu/abs/1995ApJ...455....7M}
}

@article{DESI:2025zgx,
    author = "Abdul Karim, M. and others",
    collaboration = "DESI",
    title = "{DESI DR2 results. II. Measurements of baryon acoustic oscillations and cosmological constraints}",
    eprint = "2503.14738",
    archivePrefix = "arXiv",
    primaryClass = "astro-ph.CO",
    reportNumber = "FERMILAB-PUB-25-0169-PPD",
    doi = "10.1103/tr6y-kpc6",
    journal = "Phys. Rev. D",
    volume = "112",
    number = "8",
    pages = "083515",
    year = "2025"
}

@article{Scolnic:2021amr,
    author = "Scolnic, Dan and others",
    title = "{The Pantheon+ Analysis: The Full Data Set and Light-curve Release}",
    eprint = "2112.03863",
    archivePrefix = "arXiv",
    primaryClass = "astro-ph.CO",
    doi = "10.3847/1538-4357/ac8b7a",
    journal = "Astrophys. J.",
    volume = "938",
    number = "2",
    pages = "113",
    year = "2022"
}

@article{Planck:2019nip,
    author = "Aghanim, N. and others",
    collaboration = "Planck",
    title = "{Planck 2018 results. V. CMB power spectra and likelihoods}",
    eprint = "1907.12875",
    archivePrefix = "arXiv",
    primaryClass = "astro-ph.CO",
    doi = "10.1051/0004-6361/201936386",
    journal = "Astron. Astrophys.",
    volume = "641",
    pages = "A5",
    year = "2020"
}

@article{SimonsObservatory:2018koc,
    author = "Ade, Peter and others",
    collaboration = "Simons Observatory",
    title = "{The Simons Observatory: Science goals and forecasts}",
    eprint = "1808.07445",
    archivePrefix = "arXiv",
    primaryClass = "astro-ph.CO",
    doi = "10.1088/1475-7516/2019/02/056",
    journal = "JCAP",
    volume = "02",
    pages = "056",
    year = "2019"
}

@article{SPT-3G:2025bzu,
    author = "Camphuis, E. and others",
    collaboration = "SPT-3G",
    title = "{SPT-3G D1: CMB temperature and polarization power spectra and cosmology from 2019 and 2020 observations of the SPT-3G main field}",
    eprint = "2506.20707",
    archivePrefix = "arXiv",
    primaryClass = "astro-ph.CO",
    reportNumber = "FERMILAB-PUB-25-0144-PPD",
    doi = "10.1103/7wt3-9v2y",
    journal = "Phys. Rev. D",
    volume = "113",
    number = "8",
    pages = "083504",
    year = "2026"
}

@article{AtacamaCosmologyTelescope:2025blo,
    author = "Louis, Thibaut and others",
    collaboration = "Atacama Cosmology Telescope",
    title = "{The Atacama Cosmology Telescope: DR6 power spectra, likelihoods and {\ensuremath{\Lambda}}CDM parameters}",
    eprint = "2503.14452",
    archivePrefix = "arXiv",
    primaryClass = "astro-ph.CO",
    reportNumber = "FERMILAB-PUB-25-0071-PPD",
    doi = "10.1088/1475-7516/2025/11/062",
    journal = "JCAP",
    volume = "11",
    pages = "062",
    year = "2025"
}

@book{CMB-S4:2016ple,
    author = "Abazajian, Kevork N. and others",
    collaboration = "CMB-S4",
    title = "{CMB-S4 Science Book, First Edition}",
    eprint = "1610.02743",
    archivePrefix = "arXiv",
    primaryClass = "astro-ph.CO",
    reportNumber = "FERMILAB-FN-1024-A-AE",
    doi = "10.2172/1352047",
    month = "10",
    year = "2016"
}

@article{Cortes:2025joz,
    author = "Cort{\^e}s, Marina and Liddle, Andrew R.",
    title = "{On DESI's DR2 exclusion of LCDM}",
    eprint = "2504.15336",
    archivePrefix = "arXiv",
    primaryClass = "astro-ph.CO",
    doi = "10.1093/mnrasl/slaf108",
    journal = "Mon. Not. Roy. Astron. Soc.",
    volume = "544",
    pages = "L121--L125",
    year = "2025"
}

@article{Planck:2018vyg,
    author = "Aghanim, N. and others",
    collaboration = "Planck",
    title = "{Planck 2018 results. VI. Cosmological parameters}",
    eprint = "1807.06209",
    archivePrefix = "arXiv",
    primaryClass = "astro-ph.CO",
    doi = "10.1051/0004-6361/201833910",
    journal = "Astron. Astrophys.",
    volume = "641",
    pages = "A6",
    year = "2020",
    note = "[Erratum: Astron.Astrophys. 652, C4 (2021)]"
}

@ARTICLE{2006JCAP...06..026D,
       author = {{Doran}, Michael and {Robbers}, Georg},
        title = "{Early dark energy cosmologies}",
      journal = {\jcap},
         year = 2006,
        month = jun,
       volume = {2006},
       number = {6},
          eid = {026},
        pages = {026},
          doi = {10.1088/1475-7516/2006/06/026},
archivePrefix = {arXiv},
       eprint = {astro-ph/0601544},
 primaryClass = {astro-ph},
       adsurl = {https://ui.adsabs.harvard.edu/abs/2006JCAP...06..026D}
}

@ARTICLE{2012ApJ...761..152T,
       author = {{Takahashi}, Ryuichi and {Sato}, Masanori and {Nishimichi}, Takahiro and {Taruya}, Atsushi and {Oguri}, Masamune},
        title = "{Revising the Halofit Model for the Nonlinear Matter Power Spectrum}",
      journal = {\apj},
         year = 2012,
        month = dec,
       volume = {761},
       number = {2},
          eid = {152},
        pages = {152},
          doi = {10.1088/0004-637X/761/2/152},
archivePrefix = {arXiv},
       eprint = {1208.2701},
 primaryClass = {astro-ph.CO},
       adsurl = {https://ui.adsabs.harvard.edu/abs/2012ApJ...761..152T}
}

@ARTICLE{2019ApJ...883L...3L,
       author = {{Li}, Xiaolei and {Shafieloo}, Arman},
        title = "{A Simple Phenomenological Emergent Dark Energy Model can Resolve the Hubble Tension}",
      journal = {\apjl},
         year = 2019,
        month = sep,
       volume = {883},
       number = {1},
          eid = {L3},
        pages = {L3},
          doi = {10.3847/2041-8213/ab3e09},
archivePrefix = {arXiv},
       eprint = {1906.08275},
 primaryClass = {astro-ph.CO},
       adsurl = {https://ui.adsabs.harvard.edu/abs/2019ApJ...883L...3L}
}

@ARTICLE{2021CQGra..38o3001D,
       author = {{Di Valentino}, Eleonora and {Mena}, Olga and {Pan}, Supriya and {Visinelli}, Luca and {Yang}, Weiqiang and {Melchiorri}, Alessandro and {Mota}, David F. and {Riess}, Adam G. and {Silk}, Joseph},
        title = "{In the realm of the Hubble tension-a review of solutions}",
      journal = {Classical and Quantum Gravity},
         year = 2021,
        month = jul,
       volume = {38},
       number = {15},
          eid = {153001},
        pages = {153001},
          doi = {10.1088/1361-6382/ac086d},
archivePrefix = {arXiv},
       eprint = {2103.01183},
 primaryClass = {astro-ph.CO},
       adsurl = {https://ui.adsabs.harvard.edu/abs/2021CQGra..38o3001D}
}
\end{document}